# Fine and hyperfine interactions in $^{171}$YbOH and $^{173}$YbOH

Nickolas H. Pilgram, Arian Jadbabaie, Yi Zeng and Nicholas R. Hutzler

Division of Physics, Mathematics, and Astronomy

California Institute of Technology

Pasadena, California 91125

Timothy C. Steimle[†]

School of Molecular Science

Arizona State University

Tempe, Arizona 85287

[†]Corresponding author Prof. T. C. Steimle, School of Molecular Science, Arizona State University, Tempe, Arizona 85287-1604

**Abstract**

The odd isotopologues of ytterbium monohydroxide, $^{171,173}$YbOH, have been identified as promising molecules in which to measure parity (P) and time reversal (T) violating physics. Here we characterize the $\tilde{A}^2\Pi_{1/2}(0,0,0) - \tilde{X}^2\Sigma^+(0,0,0)$ band near 577 nm for these odd isotopologues. Both laser-induced fluorescence (LIF) excitation spectra of a supersonic molecular beam sample and absorption spectra of a cryogenic buffer-gas cooled sample were recorded. Additionally, a novel spectroscopic technique based on laser-enhanced chemical reactions is demonstrated and utilized in the absorption measurements. This technique is especially powerful for disentangling congested spectra. An effective Hamiltonian model is used to extract the fine and hyperfine parameters for the $\tilde{A}^2\Pi_{1/2}(0,0,0)$ and $\tilde{X}^2\Sigma^+(0,0,0)$ states. A comparison of the determined $\tilde{X}^2\Sigma^+(0,0,0)$ hyperfine parameters with recently predicted values (M. Denis, *et al.*, J. Chem. Phys. **152**, 084303 (2020), K. Gaul and R. Berger, Phys. Rev. A **101**, 012508 (2020), J. Liu *et al.*, J. Chem. Phys. **154**, 064110 (2021)) is made. The measured hyperfine parameters provide experimental confirmation of the computational methods used to compute the P,T-violating coupling constants $W_d$ and $W_M$, which correlate P,T-violating physics to P,T-violating energy shifts in the molecule. The dependence of the fine and hyperfine parameters of the $\tilde{A}^2\Pi_{1/2}(0,0,0)$ and $\tilde{X}^2\Sigma^+(0,0,0)$ states for all isotopologues of YbOH are discussed and a comparison to isoelectronic YbF is made.

## I. INTRODUCTION

Polar molecules have emerged as ideal systems to measure symmetry-violating physics in low energy, table top experiments[1–4]. The large internal fields and closely spaced opposite parity states present in molecules provide extreme sensitivity to both time reversal (T) violating and parity (P) violating physics[5,6]. With the enhanced coherence times resulting from laser cooling, precision measurements with heavy polar molecules have the potential to probe beyond Standard Model (BSM) physics at PeV scales[3,7,8]. Additionally, P-violating effects are amplified in molecules when opposite parity states, such as neighboring rotational states, are tuned to near degeneracy via an external field[9–11]. Searches for P-violation in molecules can provide both precision tests of the standard model (SM), via measurement of nuclear spin-dependent parity violation (NSD-PV)[12,13], and sensitive probes of P-violating BSM physics[14]. The odd isotopes of the linear triatomic molecule ytterbium monohydroxide, $^{171}$YbOH and $^{173}$YbOH, have been identified as promising candidates for next generation precision measurements of P- and T-violating physics. The $^{171}$YbOH isotopologue provides sensitivity to NSD-PV[15], while the $^{173}$YbOH isotopologue is sensitive to new T,P-violating hadronic BSM physics via a measurement of the $^{173}$Yb nuclear magnetic quadrupole moment (NMQM)[7,16,17]. These molecules combine advantageous parity doublets, which are absent in their diatomic analogues, with the ability to be laser cooled[7,18].

NSD-PV arises from three major sources: vector electron-axial nucleon electroweak coupling ($V_e A_n$), the nuclear anapole moment, and the combination of nuclear electroweak charge and normal hyperfine structure. The P-odd, effective NSD-PV Hamiltonian for a single molecular electronic state is[10,15]

$$H_{NSD-PV}^{eff} = \kappa W_P (\boldsymbol{S} \times \hat{I}) \cdot \hat{n} \tag{1}$$

where $\kappa$ is the measurable NSD-PV parameter encapsulating the effects of all NSD-PV sources, $W_P$ is an effective parameter quantifying the overlap of the valence electron's wavefunction with the nucleus, $\boldsymbol{S}$ is the valence electron spin, $\hat{I} = \frac{\boldsymbol{I}}{|\boldsymbol{I}|}$ where $\boldsymbol{I}$ is the nuclear spin of the $^{171}$Yb nucleus, and $\hat{n}$ is a unit vector along the molecular symmetry axis. Currently, the only non-zero measurement of NSD-PV is in atomic Cs.[19]

The coupling of new T, P-violating BSM physics to standard model particles will result in T, P-violating energy shifts in atoms and molecules.[1,3,4] The T, P-odd effective molecular Hamiltonian describing these energy shifts is:[20]

$$H_{T,P}^{eff} = W_d d_e \boldsymbol{S} \cdot \hat{n} + W_Q \frac{Q}{I} \boldsymbol{I} \cdot \hat{n} - \frac{W_M M}{2I(2I-1)} \boldsymbol{S}\hat{T}\hat{n}, \qquad (2)$$

where $d_e$ is the electron electric dipole moment (eEDM), $Q$ is the nuclear Schiff moment (NSM), $M$ is the NMQM, $W_d$, $W_Q$, and $W_M$ are coupling constants parameterizing the sensitivity of the molecule to the different T,P-violating sources, and $\hat{T}$ is a rank 2 tensor that relates the NMQM shift to the nuclear spin orientation. Note the NMQM term is only nonzero for nuclei with $I > \frac{1}{2}$. A measurement of a non-zero eEDM, NSM, or NMQM within the projected sensitivity of contemporary experiments would be a confirmation of new T,P-violating BSM physics, while a null measurement places bounds that constrain potential BSM theories. The current limit on the eEDM[5] is $|d_e| < 1.1 \times 10^{-29}$e cm, and several upcoming molecular experiments aim to improve this limit by an order of magnitude or more.[8,18,21,22] The most stringent limit on an NSM results from measurements involving $^{199}$Hg[23]. Several other NSM experiments in other atomic[24–27] and molecular[28,29] systems are also underway. The best bounds on a NMQM are obtained from measurements of atomic Cs[30]. NMQMs are enhanced by collective effects in deformed nuclei[20,31] and the large quadrupole deformation of the $^{173}$Yb nucleus indicates that it should have an enhanced NMQM. A measurement of a NMQM in $^{173}$YbOH would complement eEDM searches as the former probes BSM physics in the hadronic sector, as opposed to the leptonic sector.[20,32,33] Furthermore, a NMQM measurement would complement NSM searches, since a measurement of both a NMQM and NSM in several different systems will allow the exact source of the hadronic BSM physics to be pinpointed.

Polyatomic molecules offer several distinct advantages over diatomic molecules for precision measurements[4,7]. First, linear triatomic molecules generically have parity doublets, known as *l*-doublets, in their excited bending modes. In YbOH and other similar molecules, the *l*-doublet splitting is ~10 MHz.[7] In the case of T,P-violation measurements, *l*-doublets enable full polarization in modest electric fields, <1 kV/cm[7], and act as internal comagnetometer states, allowing for robust control over systematic errors via reversal of the T,P-violating energy shift without switching external laboratory fields. In the case of a NSD-PV measurement, the opposite parity *l*-doublet states can be tuned to near degeneracy with modest magnetic fields, ~1-10 mT,[15] both increasing sensitivity and reducing the systematics associated with large B fields and field reversals. The electronic structure of the molecule can then be leveraged for laser cooling (if the molecule has an electronic structure amendable to laser cooling, such as YbOH) without conflicting with the internal structure needed for parity doublets. Precision measurements utilizing laser cooled and trapped molecules could increase measurement sensitivity by orders of magnitude compared to beam experiments.[7] Laser cooling of the $^{174}$YbOH isotopologue has already been demonstrated[18], and could be extended to the odd isotopologues by taking hyperfine structure into consideration.

Here, we report on the characterization of the $\tilde{A}^2\Pi_{1/2}(0,0,0) - \tilde{X}^2\Sigma_{1/2}(0,0,0)$ transition of $^{171}$YbOH and $^{173}$YbOH, which for convenience will be designated as $0_0^0\ \tilde{A}\ ^2\Pi_{1/2}\text{-}\tilde{X}\ ^2\Sigma^+$. Our study includes both excitation spectroscopy using laser induced fluorescence (LIF) detection of a cold supersonic molecular beam sample and laser absorption measurements of a cryogenic buffer-gas cooled (CBGC) sample. The derived molecular parameters are necessary for the implementation of both NSD-PV and NMQM measurements as well as for laser cooling. In order to extract the NSD-PV and NMQM parameters, $\kappa$ and $M$, from P- or T-violating molecular energy shifts, the values of the coupling constants $W_P$ or $W_M$ must be known. These coupling constants cannot be measured experimentally and instead must be calculated via *ab initio* or semi-empirical methods. Comparison of measured hyperfine parameters and calculated ones provide a rigorous test of the computational methods used to calculate the T,P-violating coupling constants, as both sets of parameters probe the nature of the valence electron in the vicinity of the nucleus. Finally, we demonstrate a novel spectroscopic technique utilizing laser-induced chemical reactions[34] to both amplify the signal from the desired isotopologue and disentangle the respective isotopologue's spectrum from that of other, more abundant, overlapping isotopologues.

A review of experimental measurements and theoretical predictions for YbOH can be found in the manuscript describing high-resolution optical analysis performed on the even isotopologues[35]. More recent experimental work includes the aforementioned demonstration of Sisyphus and Doppler laser cooling of $^{174}$YbOH[18] as well as the determination of fluorescence branching ratios, radiative lifetimes and transition moments[36]. Recently reported theoretical studies include calculations of the molecular NMQM sensitivity coefficient, $W_M$, of Eq. 2[16,17]. As part of these studies, the ground state magnetic hyperfine parameters, $A_\parallel$ for $^{171}$YbOH and $^{173}$YbOH, and the nuclear electric quadrupole coupling parameter, $e^2q_0Q$ for $^{173}$YbOH, were predicted.

**EXPERIMENTAL**

Initial experiments were performed at ASU using a molecular beam LIF spectrometer similar to that used in previous high-resolution optical studies of $^{172,174}$YbOH.[35,37] Briefly, YbOH is produced by laser ablating (532 nm, ~10 mJ/pulse, 20 Hz) a Yb rod in the presence of a methanol/argon supersonic expansion. The resulting beam is skimmed to produce a well collimated beam with a temporal pulse width of ~40 $\mu s$ in the detection region. The molecular beam is probed by an unfocused (~5 mm), low power (~5 mW), single frequency cw-dye laser approximately 0.5 m downstream. It is estimated that the laser probes approximately $1\times10^9$ YbOH molecules in each molecular beam pulse. The resulting on-resonance LIF signal was viewed through a 580 ± 10 nm bandpass filter, detected by a photomultiplier tube (PMT), and processed using gated photon counting. Typically, the photon counts from 35 ablation pulses at each excitation laser frequency are summed. The absolute excitation wavelength is determined by co-recording a sub-Doppler $I_2$ spectrum[38], and the relative wavelength is measured by co-recording the transmission of an actively stabilized etalon (free spectral range of 751.393 MHz).

Subsequent high-resolution spectra were recorded at Caltech using absorption spectroscopy. The apparatus was nearly identical to that used in the initial demonstration of laser-induced chemical enhancement of YbOH production[34]. In summary, cold YbOH molecules are produced via cryogenic buffer-gas cooling, which is described in detail elsewhere.[39–43] The molecules are produced inside a copper, cryogenic buffer-gas cell cooled to ~4 K, with an internal cylindrical bore of 12.7 mm, and a length of ~83 mm. Helium buffer-gas is introduced into the cell by a 3.2 mm diameter copper tube. The helium then passes through a diffuser 3.2 mm from the gas inlet and exits the cell through a 5 mm aperture on the other end of the cell. All measurements presented here were performed inside the buffer-gas cell, as opposed to in the extracted beam. YbOH molecules are created by ablating (532 nm, ~30 mJ/pulse, ~5 Hz) solid pressed targets of Yb powder mixed with either $Yb(OH)_3$ or $Te(OH)_6$ powders. To make the $Yb+Yb(OH)_3$ targets, the powders were mixed to obtain a 1:1 stoichiometric ratio of Yb to OH, then ground in a mortar and pestle, passed through a 230 mesh sieve, mixed with 4% polyethylene glycol binder (PEG8000) binder by weight, and pressed in an 8 mm die at ~10 MPa for ~15 minutes. A similar procedure was used to produce the $Yb+Te(OH)_6$ targets.

To measure the absorption spectra, three cw-laser beams are passed through the spectroscopy window: the primary tunable absorption spectroscopy beam (1 mm diameter, ~30 $\mu$W), the normalization laser (1 mm diameter, ~40 $\mu$W) used to monitor the shot-to-shot fluctuations in YbOH production, and the chemical enhancement laser (3 mm diameter, ~300 mW), which increases the molecular yield by exciting atomic Yb to the metastable $^3P_1$ state, thereby boosting chemical reactions rates between Yb atoms and OH-containing ablation products[34].

The normalization laser is fixed to either the $^OP_{12}(2)$ or $^RR_{11}(2)$ line of the $0_0^0$ $\tilde{A}\,^2\Pi_{1/2}$-$\tilde{X}\,^2\Sigma^+$ transition of $^{172}$YbOH at 17322.1732 cm$^{-1}$ and 17327.0747 cm$^{-1}$, respectively[35]. The frequency of the enhancement laser is fixed to the $^3P_1$-$^1S_0$ transition of the desired Yb isotope. Specifically, the $^{174}$Yb(17992.0003 cm$^{-1}$), $^{176}$Yb(17991.9685 cm$^{-1}$), $F'' = 1/2 \rightarrow F' = 1/2$ $^{171}$Yb (17991.9292 cm$^{-1}$) and the $F'' = 5/2 \rightarrow F' = 7/2$ $^{173}$Yb (17991.9207 cm$^{-1}$) transitions[44] were used. The fixed lasers are locked to a stabilized HeNe laser via a scanning transfer cavity and active feedback (~5 MHz resolution). The primary absorption laser is continuously scanned in frequency and the resulting absorption is detected with a photodiode. The absolute frequency of the primary absorption laser is monitored by a digital wavemeter and the relative frequency is tracked in a separate transfer cavity with respect to a stabilized HeNe laser (~7 MHz resolution). The absolute transition frequencies are calibrated using known $^{172}$YbOH and $^{174}$YbOH spectral lines[35]. The light of the enhancement laser is switched on and off using a mechanical shutter so that both enhanced and unenhanced spectra of the desired isotopologue can be measured in successive shots.

The chemical enhancement is utilized as a spectroscopic tool to disentangle the complicated isotopologue structure. This technique is illustrated for $^{171}$YbOH in Figure 1 and described below. The measured optical depth (OD) of the primary probe beam is integrated over the duration of the molecular pulse, typically ~3 ms. This integrated OD is then normalized by the integrated OD of the normalization probe. The normalized signal at a single laser frequency without enhancement light is $S_{UE} = S_{171} + S_B$, where $S_{171}$ is the measured integrated OD for a specific transition of $^{171}$YbOH, and $S_B$ is the background integrated OD from all other overlapping YbOH isotopologues or other molecules. The measured normalized and integrated OD with the enhancement light on and tuned to the $^{171}$Yb atomic resonance is $S_E = E S_{171} + S_B$. Here, $E$ is the enhancement factor, $N_E/N_0$, the ratio of the number of molecules produced with the enhancement laser on, $N_E$, to the number produced with the enhancement laser off, $N_0$. Taking the difference between the enhanced and unenhanced signals $S_E - S_{UE} = (E - 1)S_{171}$ results in the spectrum from only the $^{171}$YbOH isotopologue. It is observed that $E$ in the cell is typically ~4-8. This technique generalizes to the other isotopologues of YbOH and other molecules as well.

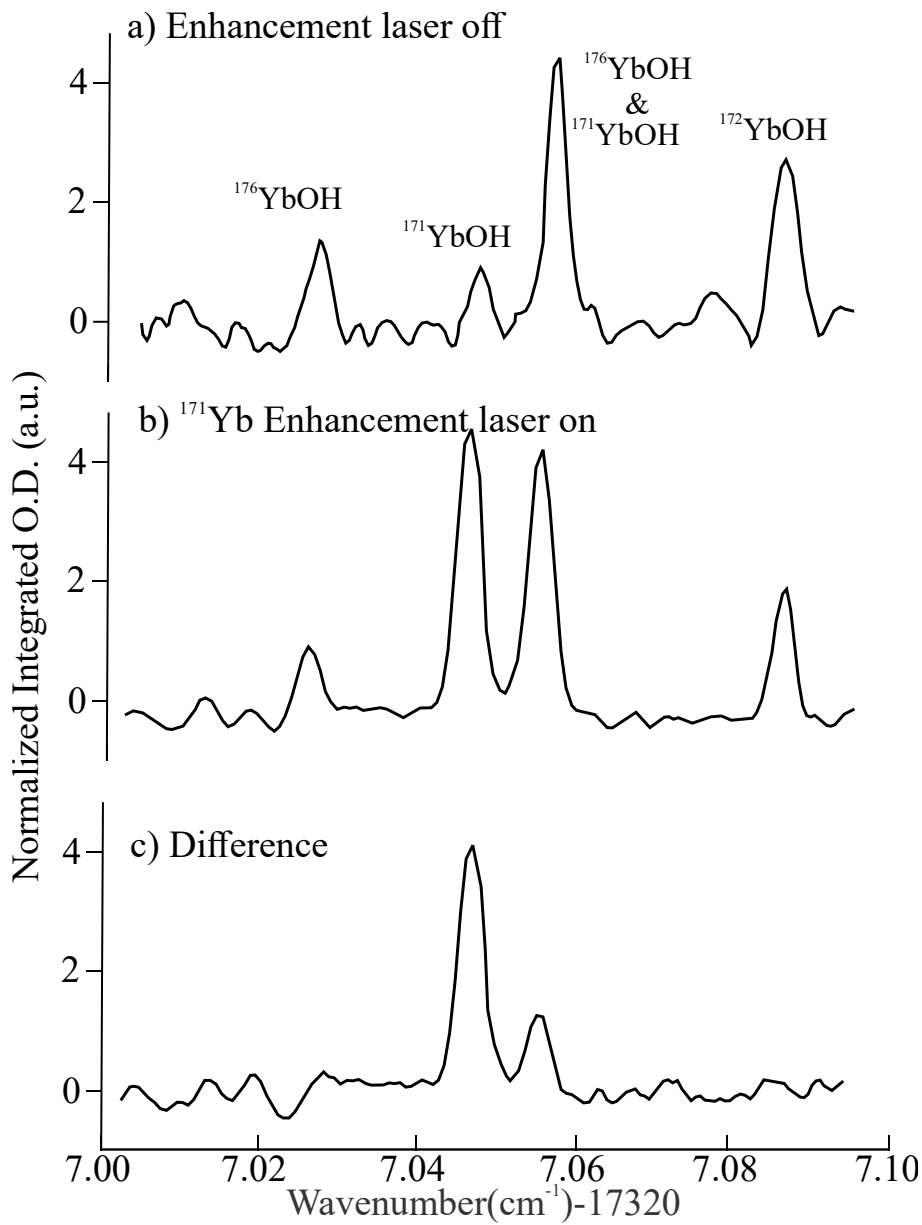

**Figure 1.** Extraction of the $^{171}$YbOH spectrum in the region of the even isotopologue $^RR_{11}(2)$ lines using chemical enhancement. These data are taken from a cryogenic buffer-gas cooled sample. **a.)** Spectrum with no chemical enhancement. The lines of each isotopologue are indicated. The $^{174}$YbOH $^RR_{11}(2)$ line is overlapped with a much weaker $^{171}$YbOH line. **b.)** Spectrum with $^{171}$YbOH chemical enhancement. **c.)** Difference of the enhanced and non-enhanced spectrum (a-b), this spectrum is purely from the $^{171}$YbOH isotopologue.

**OBSERVATION**

A general description of the $0_0^0$ $\tilde{A}\,^2\Pi_{1/2}$-$\tilde{X}\,^2\Sigma^+$ band can be found in the previous report on the analysis of the field-free, Stark and Zeeman spectroscopy of $^{172,174}$YbOH.[35] Here we focus on the odd isotopologues. There are seven naturally occurring isotopes of Yb in moderate abundance: $^{168}$Yb (0.1%), $^{170}$Yb (3.0%), $^{171}$Yb (14.3%), $^{172}$Yb (21.8%), $^{173}$Yb (16.1%), $^{174}$Yb (31.8%), and $^{176}$Yb (12.8%). The $^{171}$YbOH and $^{173}$YbOH spectra are more complex than those for $^{172,174}$YbOH, primarily due to the large $^{171}$Yb(I=1/2, $\mu$ = +0.49367$\mu_N$) and $^{173}$Yb(I=5/2, $\mu$ = -0.67989$\mu_N$,) magnetic hyperfine interaction and, in the case of $^{173}$YbOH, the large nuclear electric quadrupole hyperfine interaction ($Q$=280.0 ± 4.0 fm$^2$)[45]. The previously described branch designation used for the even and odd isotopologues of the (0,0) $A\,^2\Pi_{1/2}$-$X\,^2\Sigma^+$ band of YbF [46] will also be used for the $0_0^0$ $\tilde{A}\,^2\Pi_{1/2}$-$\tilde{X}\,^2\Sigma^+$ band YbOH. The even isotopologues exhibit six branches, labelled $^PP_{11}$, $^QQ_{11}$, $^RR_{11}$, $^PQ_{12}$, $^OP_{12}$ and $^QR_{12}$ following the $^{\Delta N}\Delta J_{F_i'F_i''}(N'')$ designation expected for a $^2\Pi_{1/2}$(Hund's case ($a$))- $^2\Sigma$(Hund's case ($b$)) band. The low rotational levels of the $\tilde{X}\,^2\Sigma^+$ state for the odd isotopologues exhibit a Hund's case ($b_{\beta S}$) energy level pattern due to the large $^{171}$Yb(I=1/2) and $^{173}$Yb(I=5/2) magnetic hyperfine interaction, where the electron spin angular momentum, **S**, is coupled to the $^{171}$Yb(I=1/2) or $^{173}$Yb(I=5/2) nuclear spin angular momentum, **I**, to give an approximately good intermediate quantum number $G$. The angular momentum **G** is coupled to the rotational angular momentum, **N**, to give the intermediate angular momentum **F**$_1$, which is then coupled to the proton nuclear spin angular momentum **I$_2$**(=1/2) to give the total angular momentum **F**. The corresponding coupling limit wavefunction, $\left|(SI(Yb))G(GN)F_1(F_1I(H))F\right\rangle$, is useful for describing the low-rotational energy levels of the $\tilde{X}\,^2\Sigma^+(0,0,0)$ state. The Yb and H hyperfine interactions in the $\tilde{A}\,^2\Pi_{1/2}$ state are very small compared to the rotational and Λ-doubling spacing, and the energy level pattern is that of a molecule near a sequentially coupled Hund's case $(a_{\beta J})$ limit. The corresponding coupling limit wavefunction, $|\eta\Lambda\rangle|S\Sigma\rangle|J\Omega(JI(Yb))F_1(F_1I(H))F\rangle$, is useful in describing the low-rotational energy levels of the $\tilde{A}\,^2\Pi_{1/2}(0,0,0)$ state. With the exception of broadening of the lowest rotational branch features of the molecular beam spectra, there was no evidence of proton hyperfine splitting. The six $^2\Pi_{1/2}$(Hund's case ($a$)) - $^2\Sigma$(Hund's case ($b$)) branches of the even isotopologues split and regroup into eight branches labeled $^OP_{1G}$, $^PP_{1G}+\,^PQ_{1G}$, $^QQ_{1G}+\,^QR_{1G}$, and $^RR_{1G}$, appropriate for a $^2\Pi_{1/2}$(case $(a_{\beta J})$)- $^2\Sigma$( case $(b_{\beta S})$) band with $G=0$ and 1 for $^{171}$YbOH and $G=2$ and 3 for $^{173}$YbOH. The lines of the $^OP_{1G}$ and $^RR_{1G}$ branches (odd isotopologues) and the $^OP_{12}$ and $^RR_{11}$ branches (even isotopologues) form progressions in $N''$ with adjacent members separated by $\sim 4B''$ extending to the red and blue, respectively, and are relatively unblended. The $0_0^0$ $\tilde{A}\,^2\Pi_{1/2}$- $\tilde{X}\,^2\Sigma^+$ band exhibits a blue degraded head formed by low- $N''$ features of the $^PQ_{12}$, $^PP_{11}$ (even isotopologues) and $^PP_{1G}+\,^PQ_{1G}$ (odd isotopologues) branches. Isotopic spectral shifts are very small because the potential energy surfaces for the $\tilde{X}\,^2\Sigma^+(0,0,0)$ and $\tilde{A}\,^2\Pi_{1/2}(0,0,0)$ states are very similar, causing the $^PQ_{12}$, $^PP_{11}$, $^PP_{11}$, $^QQ_{11}$, and $^QR_{12}$ branches of the even isotopologues and $^PP_{1G}+\,^PQ_{1G}$, and $^QQ_{1G}+\,^QR_{1G}$ branches of the odd isotopes to be severely overlapped.

The observed and calculated LIF molecular beam spectra in the region of the $^OP_{12}$ (3) even isotopologues branch features and the $^OP_{1G}$ (3) odd isotopologues branch features are presented in Figure 2. The observed spectrum in Figure 2 is similar to that of the (0,0) $A\,^2\Pi_{1/2}$- $X\,^2\Sigma^+$ band of YbF (Ref.[46]) with the exception that a small splitting (~50 MHz) due to $^{19}$F(I=1/2) was observed in YbF, whereas the smaller H(I=1/2) doubling in YbOH is not fully resolved. The predicted spectrum was obtained assuming a 15 K rotational temperature, 30 MHz full width at half maximum (FWHM) Lorentzian line shape and the optimized parameters (see below). The two components of the $^OP_{1G}$ (3) transition are widely spaced and of the opposite order because the spacing between the $X\,^2\Sigma^+$ (0,0,0) $G$=3 and $G$=2 groups of levels of $^{173}$YbOH is approximately -5660 MHz ($\sim 3\,b_F''$) and that between the $G$=1 and $G$=0 levels of $^{171}$YbOH is approximately +6750 MHz ($\sim b_F''$). The spectral patterns of the $G$=3 and $G$=2 groups of $^{173}$YbOH are irregular because of the large nuclear electric quadrupole interaction ($e^2q_0Q$) in both the $\tilde{X}\,^2\Sigma^+(0,0,0)$ and $\tilde{A}\,^2\Pi_{1/2}(0,0,0)$ states.

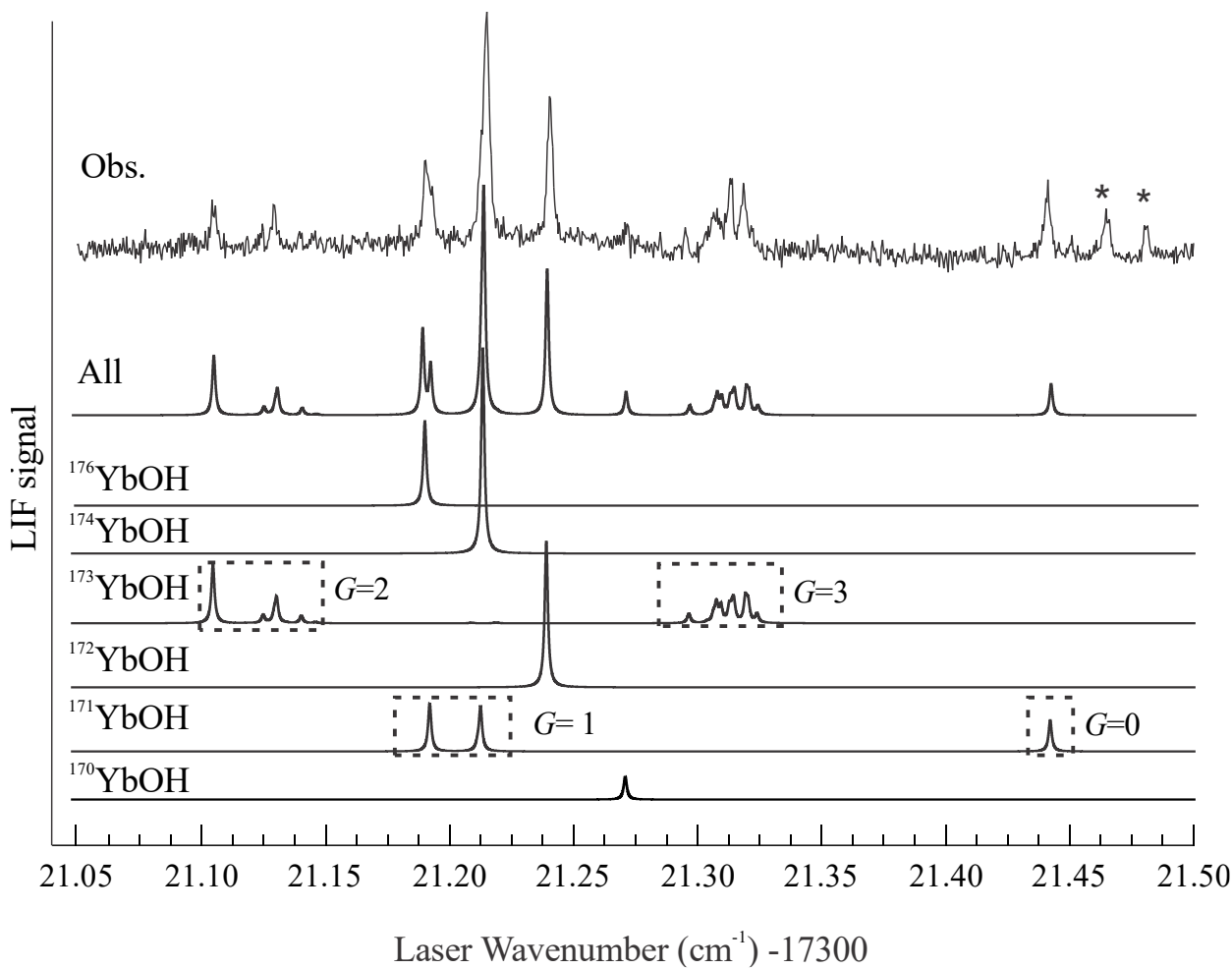

**Figure 2.** The observed and calculated LIF spectra in the region of the $^OP_{12}$(3) even isotopologues branch features and the $^OP_{1G}$(3) odd isotopologues of the $0_0^0$ $A\,^2\Pi_{1/2}$ - $X\,^2\Sigma^+$ band of YbOH. The predicted spectra for the even isotopologues were obtained using the optimized parameters of Ref.[35] and those for $^{171}$YbOH and $^{173}$YbOH obtained using optimized parameters of Table I. A rotational temperature of 15 K and a Lorentzian fullwidth at half maximin (FWHM) linewidth of 30 MHz were used. The features marked "*" are unidentified.

The laser-induced fluorescence spectrum of a molecular beam sample in the highly congested bandhead region (i.e. between 17,323.50 and 17,323.85 cm$^{-1}$) is presented in Figure 1 of Ref.[35] The bandhead region is dominated by low-rotational $^PP_{11}$, $^QQ_{11}$, $^PQ_{12}$ and $^QR_{12}$ branch features of the more abundant even isotopologues, which makes assignment of the $^PP_{1G} + {}^PQ_{1G}$ and $^QQ_{1G} + {}^QR_{1G}$ branch features of the less abundant odd isotopologues of the molecular beam sample extremely difficult. Recording the chemically enhanced absorption spectrum of a CBGC sample in the bandhead region was critical to the assignment and subsequent analysis. The observed and predicted high-resolution absorption spectra in the bandhead region of a CBGC sample is presented in Figure 3, both with (Panel b) and without (Panel a) $^{173}$YbOH chemical enhancement induced by the $^{173}$Yb atomic excitation. There is approximately a factor of four increase in signal upon atomic excitation. The difference with and without chemical enhancement (Panel a-Panel b) is presented in Panel c of Figure 3. The predicted spectrum was obtained assuming a 5 K rotational temperature, 90 MHz FWHM Lorentzian lineshape and the optimized parameters (see below). There is no evidence of proton hyperfine splitting at this spectral resolution. The CBGC spectra are greatly simplified relative to the supersonic molecular beam spectra due to both the isotopic selectivity, which is evident form Figure 3 (Panel a versus Panel c), and the lower rotational temperature (~5 K versus ~20K). Even so, the spectral features of Figure 3 (Panel c) are blends of many transitions (see below). For example, the feature marked "A" in Figure 3 is predominantly a blend of the $F_1'' = 3 \rightarrow F_1' = 3$ component of the $^PP_{12} + {}^PQ_{12}(1)$ branches and the $F_1'' = 4 \rightarrow F_1' = 4$ component of the $^PP_{12} + {}^PQ_{12}(2)$ branch features of $^{173}$YbOH. The feature marked "B" is predominantly a blend of the $^PP_{12} + {}^PQ_{12}(3)$ ($F_1'' = 5 \rightarrow F_1' = 5$), $^PP_{12} + {}^PQ_{12}(1)$ ($F_1'' = 2 \rightarrow F_1' = 3$) and $^PP_{12} + {}^PQ_{12}(3)$ ($F_1'' = 4 \rightarrow F_1' = 4$) transitions.

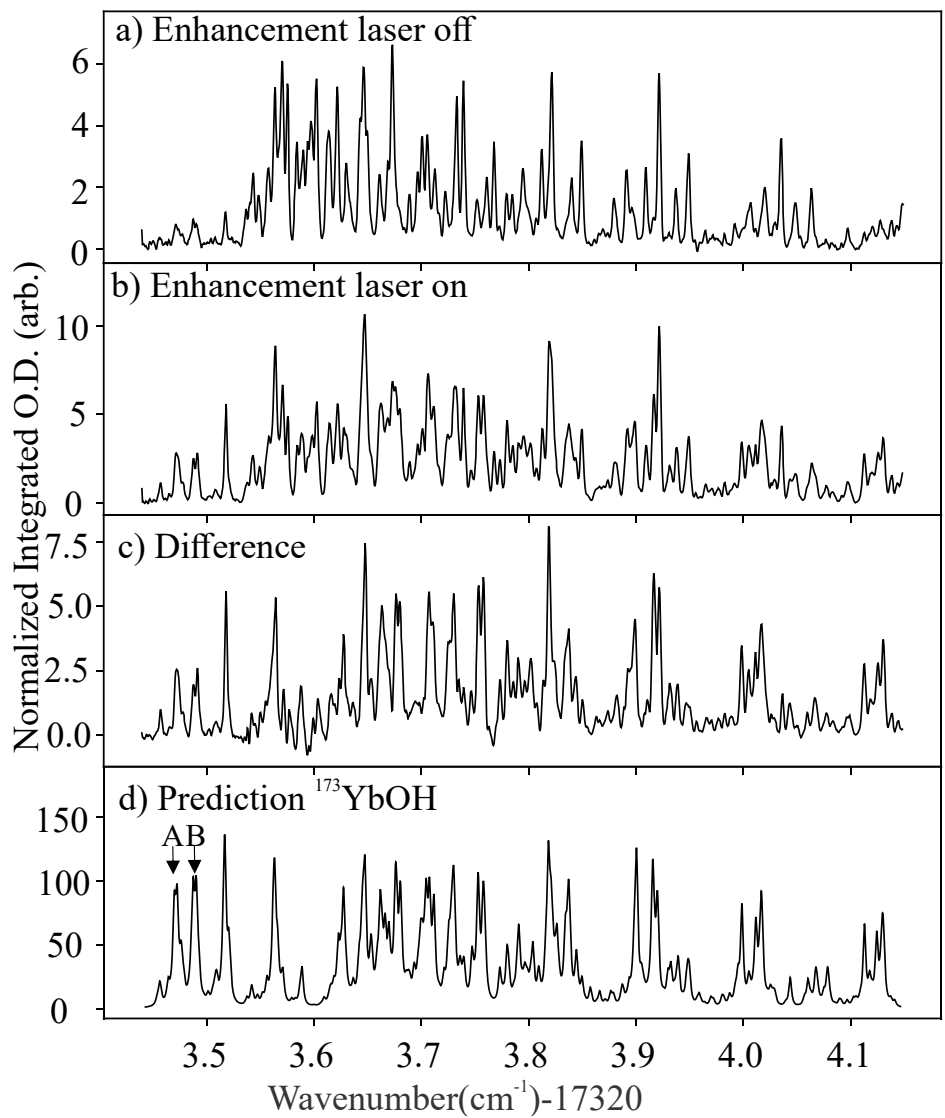

**Figure 3.** High-resolution absorption spectra in the bandhead region of the $0_0^0$ $A\,^2\Pi_{1/2}$ - $X\,^2\Sigma^+$ band of YbOH of a cryogenic buffer-gas cooled sample: **a)** Spectrum with no chemical enhancement; **b)** spectrum recorded when the $F'' = 5/2 \rightarrow F' = 7/2$ component of the $^3P_1$-$^1S_0$ transition of $^{173}$Yb (17991.9207 cm$^{-1}$) is excited; **c)** difference of enhanced and unenhanced spectra (**a-b**); **d)** prediction of the $^{173}$YbOH absorption spectrum using the optimized parameters given in Table I, a rotational temperature of 5 K and FWHM linewidth of 90 MHz. Units of **d** y-axis are arbitrary.

The observed and predicted high-resolution absorption spectra in the bandhead region of the CBGC sample is presented in Figure 4, both with (Panel b) and without (Panel a) $^{171}$YbOH chemical enhancement induced by the $^{171}$Yb atomic excitation. The spectrum obtained by subtracting the signals with and without chemical enhancement induced by the $^{171}$Yb atomic excitation is presented in Panel c. The predicted spectrum was obtained assuming a 5 K rotational temperature, 90 MHz FWHM Lorentzian lineshape and the optimized parameters (see below). The head at 17323.55 cm$^{-1}$ is an unresolved blend of the $^{P}P_{12} + {}^{P}Q_{12}(1)$, $^{P}P_{12} + {}^{P}Q_{12}(2)$, and $^{P}P_{12} + {}^{P}Q_{12}(3)$ branch features (see below). Unlike the $^{173}$YbOH spectrum of Figure 3, there are numerous $^{171}$YbOH spectral features that are unblended. For example, the feature marked "A" in Figure 4 is the $F_1'' = 1 \rightarrow F_1' = 1$ component of $^{Q}Q_{11} + {}^{Q}R_{11}(1)$, "B" the $F_1'' = 5 \rightarrow F_1' = 5$ component of $^{P}P_{11} + {}^{P}Q_{11}(5)$, and "C" the $F_1'' = 2 \rightarrow F_1' = 2$ component of $^{Q}Q_{11} + {}^{Q}R_{11}(2)$ (see below).

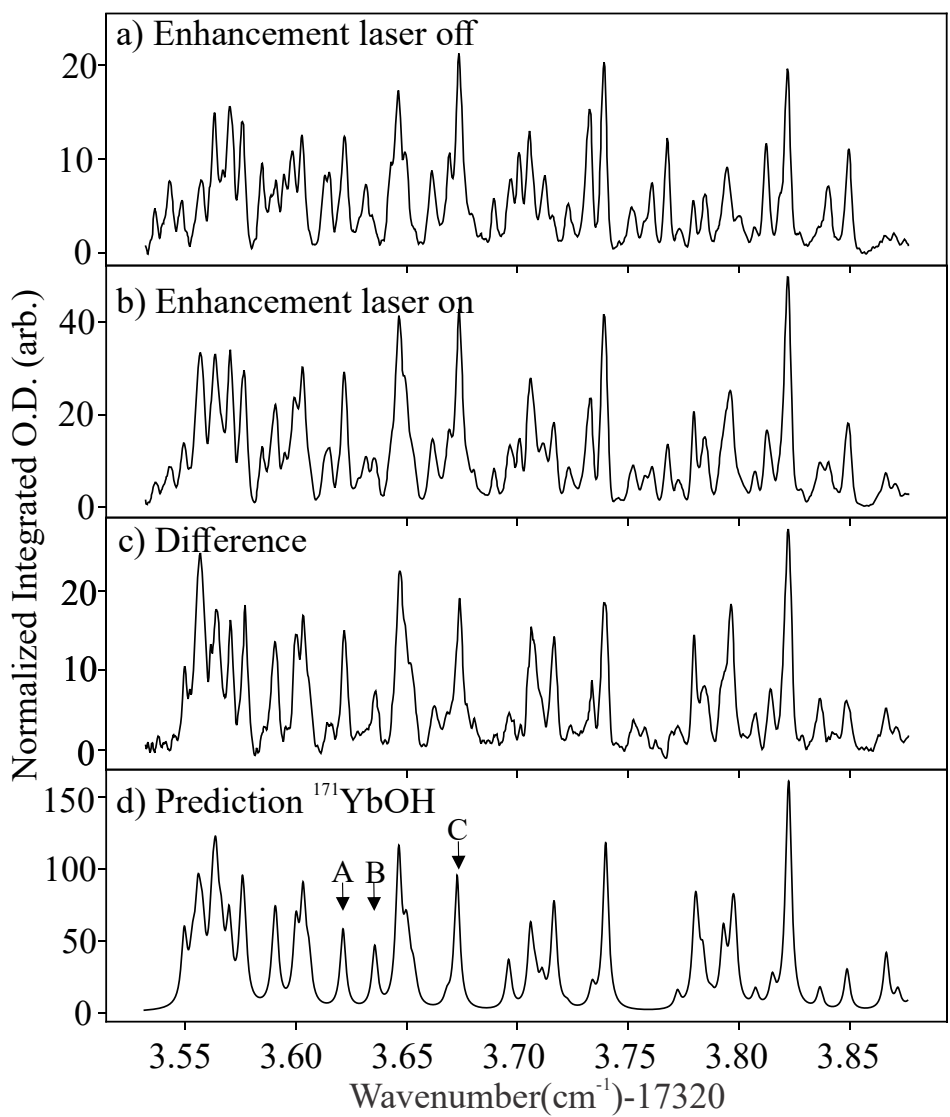

**Figure 4.** High-resolution absorption spectra in the bandhead region of the $0_0^0$ $A\,^2\Pi_{1/2}$ - $X\,^2\Sigma^+$ band of YbOH of a cryogenic buffer-gas cooled sample: **a)** Spectrum with no chemical enhancement; **b)** spectrum recorded when the $F'' = 1/2 \rightarrow F' = 1/2$ component of the $^3P_1$-$^1S_0$ transition of $^{171}$Yb (17991.9292 cm$^{-1}$) is excited; **c)** difference of enhanced and unenhanced spectra (**a-b**); **d)** prediction of the $^{171}$YbOH absorption spectrum using the optimized parameters given in Table I, a rotational temperature of 5 K and FWHM linewidth of 90 MHz. Units of **d** y-axis are arbitrary.

Although the $^RR_{1G}$ branch features of the odd isotopologues are less blended than the $^PP_{1G}$ + $^PQ_{1G}$ and $^QQ_{1G}$+ $^QR_{1G}$ branch features, recording the absorption spectrum of the CBGC sample was still critical for spectral disentanglement and assignment of the molecular beam LIF spectrum. This is illustrated in Figure 5 where the observed and predicted molecular beam LIF (left side) and CBGC absorption (right side) spectra in the region of the $^RR_{11}(2)$ (even isotopologues) and $^RR_{1G}(2)$ (odd isotopologues) lines are presented. Predicted molecular beam spectra assumed a rotational temperature of 20K and FWHM linewidth of 30 MHz, while the CBGC spectra assumed a 5 K and 90MHz and both used the optimized parameters (see below). The CBGC absorption spectrum recorded without the atomic excitation is presented in Panel a. The CBGC absorption spectra recorded with the atomic excitation laser tuned to the $^3P_1$-$^1S_0$ transitions of $^{176}$Yb, $^{174}$Yb and $^{171}$Yb are presented Panels b, c and d, respectively. The $^{171}$YbOH absorption spectrum in Panel e was obtained by taking the difference of "a" and "d", revealing a spectral feature that is obscured in the higher resolution molecular beam LIF spectrum by the much more intense $^RR_{11}(2)$ line of $^{174}$YbOH.

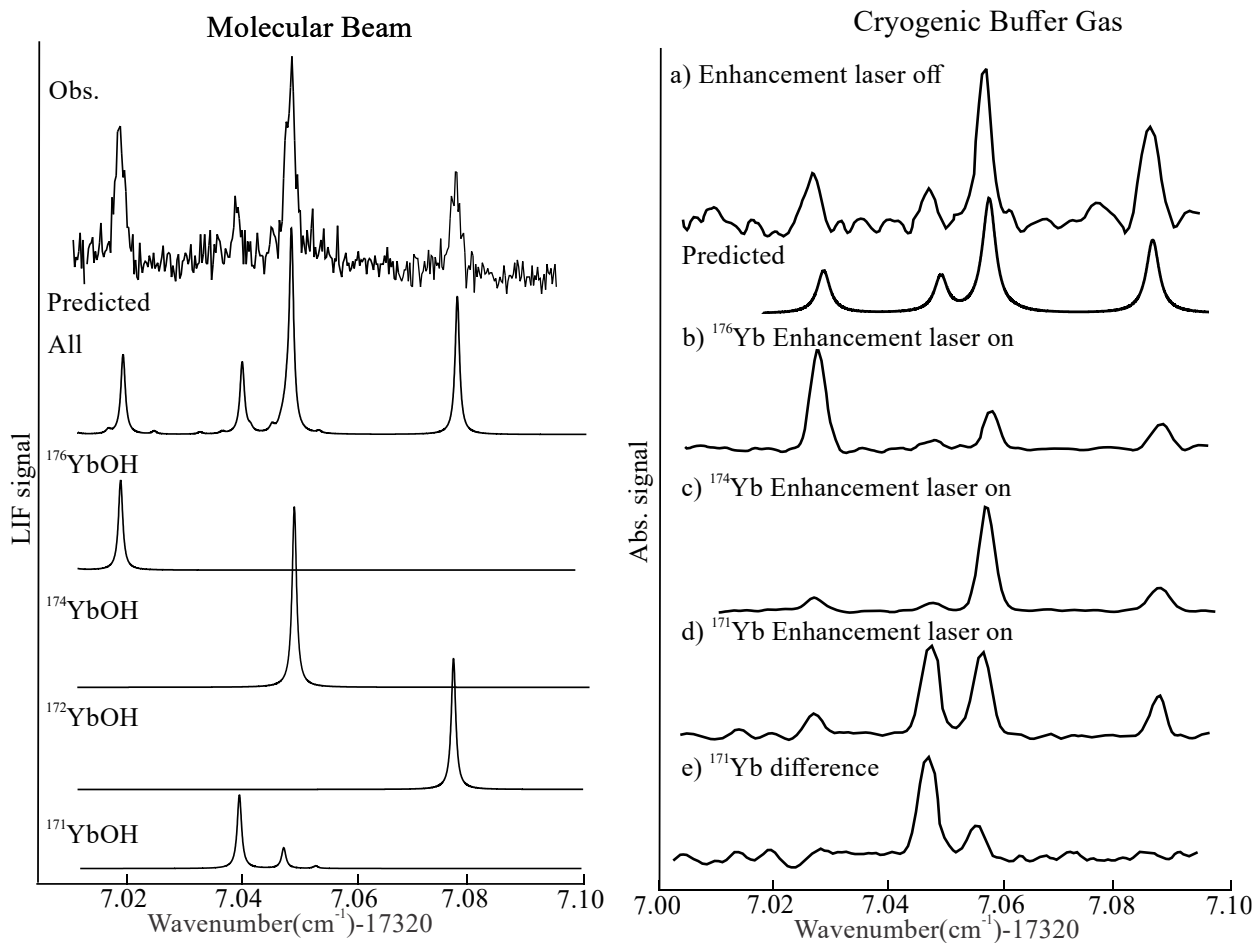

**Figure 5:** Observed and predicted molecular beam (MB) LIF (left side) and cryogenic buffer-gas cooled (CBGC) absorption (right side) spectra in the region of the $^RR_{11}(2)$ line of the $0_0^0$ $A\,^2\Pi_{1/2}$ - $X\,^2\Sigma^+$ band of YbOH. Predicted MB spectra assumed a rotational temperature of 15 K and FWHM linewidth of 30 MHz, while the CBGC spectrum used T=5 K and FWHM=90 MHz, and both used the optimized parameters of Table I. a) recorded with the atomic excitation laser blocked; b) recorded with the atomic excitation laser tuned to the $^3P_1$-$^1S_0$ transition of $^{176}$Yb (17991.9685 cm$^{-1}$); c) recorded with the atomic excitation laser tuned to the $^3P_1$-$^1S_0$ transition of $^{174}$Yb (17992.0003 cm$^{-1}$); d) recorded with the atomic excitation laser tuned to the $F'' = 1/2 \rightarrow F' = 1/2$ component of the $^3P_1$-$^1S_0$ transition of $^{171}$Yb (17991.9292 cm$^{-1}$) e) difference of "a" and "d", showing direct, model-free isolation of $^{171}$YbOH signal from other isotopologues.

The precisely measured transition wavenumber for 94 spectral features of the $0_0^0$ $\tilde{A}\,^2\Pi_{1/2}$-$\tilde{X}\,^2\Sigma^+$ band of $^{173}$YbOH, which are assigned to 124 transitions, are given in Table S1 of the Supplementary Material, along with the difference between the observed and calculated wavenumbers. The precisely measured transition wavenumber for 63 spectral features of the $0_0^0$ $\tilde{A}\,^2\Pi_{1/2}$-$\tilde{X}\,^2\Sigma^+$ band of $^{171}$YbOH, which are assigned to 68 transitions, are given in Table S2 of the Supplementary Material, along with the difference between the observed and calculated wavenumbers. Also presented are the quantum number assignments.

**ANALYSIS**

The analysis of the $^{171}$YbOH and $^{173}$YbOH $0_0^0$ $\tilde{A}\,^2\Pi_{1/2}$-$\tilde{X}\,^2\Sigma^+$ spectra was nearly identical to that of the $(0,0)$ $A\,^2\Pi_{1/2}$-$X\,^2\Sigma^+$ band of $^{171}$YbF(Ref. [47]) and $^{173}$YbF(Ref.[46]) and will not be described in detail here. Briefly, the observed transition wavenumbers listed in Tables S1 and S2 of the Supplementary Material were used as input into a weighted least-squares optimization routine. The molecular beam data was given a weight twice of the more extensive CBGC data due to the high spectral resolution (30 MHz FWHM *vs*. 90 MHz) and the fact that it was co-recorded with the I$_2$ wavelength calibration spectrum. The $\tilde{X}\,^2\Sigma^+(0,0,0)$ and $\tilde{A}\,^2\Pi_{1/2}(0,0,0)$ energies were obtained by diagonalizing a 24×24 $(=(2S+1)(2I_1+1)(2I_2+1))$ and 48×48 $(=2(2S+1)(2I_1+1)(2I_2+1))$ matrices for $^{173}$YbOH and 8×8 and 16×16 matrices for $^{171}$YbOH constructed in a sequentially coupled Hund's case $(a_{\beta J})$ basis set, $|\eta\Lambda\rangle|S\Sigma\rangle|J\Omega(JI(Yb))F_1(F_1I(H))F\rangle$. Note that a Hund's case $(a_{\beta J})$ basis is used for the $\tilde{X}\,^2\Sigma^+(0,0,0)$ state even though the energy level pattern is that of a molecule close to a Hund's case $(b_{\beta S})$. The $\tilde{X}\,^2\Sigma^+(0,0,0)$ states were modeled using an effective Hamiltonian[48,49] that included rotation (*B* and *D*), spin-rotation (γ), Yb and H magnetic hyperfine (Fermi contact, $b_F$, and dipolar, $c$), and $^{173}$Yb axial nuclear electric quadrupole ($e^2q_0Q$) parameters. The energy levels for the $\tilde{A}^2\Pi_{1/2}(0,0,0)$ states were modeled using an effective Hamiltonian that included the spin-orbit ($A$), rotation ($B$ and $D$), $\Lambda$-doubling ($p+2q$), Yb magnetic hyperfine ($\Lambda$-doubling type, $d$, and electronic orbital hyperfine, $a$) and $^{173}$Yb axial nuclear electric quadrupole ($e^2q_0Q$). In the Hund's case ($a_{\beta J}$) limit, the determinable combination of magnetic hyperfine parameters is $h_{1/2} \equiv a - \dfrac{b_F}{2} - \dfrac{c}{3}$ and the levels can be accurately modeled using $a\hat{I}_z\hat{L}_z$ and by fixing the Fermi contact, $b_F$, and dipolar, $c$, terms to zero. The data set, which is restricted to the $|\Omega|=1/2$ levels of the $\tilde{A}^2\Pi(0,0,0)$ state, is insensitive to the perpendicular component of the nuclear electric quadrupole interaction ($e^2q_2Q$). The effective Hamiltonians are described in more detail in the Supplementary Material.

Analyses that floated various combinations of parameters were attempted. In all cases, the small proton magnetic hyperfine parameters for the $\tilde{X}^2\Sigma^+(0,0,0)$ state were held fixed to those determined from the analysis of the $^{174}$YbOH microwave spectrum ($b_F$ = 4.80 MHz and $c$ = 2.46 MHz)[37]. The spin-orbit parameter, $A$, of the $\tilde{A}^2\Pi(0,0,0)$ state was constrained to the previously determined[50] value of 1350 cm$^{-1}$. The analysis is relatively insensitive to the value of $A$, other than the obvious linear displacement, because of the large separation between the $\tilde{A}^2\Pi_{1/2}(0,0,0)$ and $\tilde{A}^2\Pi_{3/2}(0,0,0)$ states and the relatively small rotational parameter, $B$ (i.e. the rotationally induced spin-uncoupling effect is negligible because $B/A \cong 2\times 10^{-4}$). The centrifugal correction parameters, $D$, for the $\tilde{X}^2\Sigma^+(0,0,0)$ and $\tilde{A}^2\Pi(0,0,0)$ states and the small spin-rotation parameter, $\gamma$, for the $\tilde{X}^2\Sigma^+(0,0,0)$ state were constrained to the values predicted from extrapolation of the values for $^{174}$YbOH using the expected isotopic mass dependence. In the end, the $^{173}$YbOH data set was satisfactorily fit by optimizing 10 parameters and the $^{171}$YbOH data set by optimizing 8 parameters. The standard deviation of the $^{171}$YbOH and $^{173}$YbOH fits were 27 and 25 MHz, respectively, which are commensurate with estimated weighted measurement uncertainty. The optimized parameters and associated errors are presented in Table I, along with those for the $A^2\Pi_{1/2}(v=0)$ and $X^2\Sigma^+(v=0)$ states of $^{171}$YbF(Ref. [47]) and $^{173}$YbF(Ref.[46]). The spectroscopic parameters are independently well determined with the largest correlation coefficient being between $B''$ and $B'$, which for the $^{171}$YbOH and $^{173}$YbOH fits are 0.854 and 0.960, respectively.

**Table I**. Parameters in wavenumbers (cm$^{-1}$) for the $\tilde{X}^2\Sigma^+(0,0,0)$ and $\tilde{A}^2\Pi_{1/2}(0,0,0)$ states of $^{171}$YbOH and $^{173}$YbOH and the $A^2\Pi_{1/2}(v=0)$ and $X^2\Sigma^+(v=0)$ states of $^{171}$YbF and $^{173}$YbF.

|  | Par. | $^{171}$YbOH | $^{171}$YbF[a] | $^{173}$YbOH | $^{173}$YbF[b] |
|---|---|---|---|---|---|
| $\tilde{X}^2\Sigma^+$ | $B$ | 0.245497(22) | 0.24171098(6) | 0.245211(18) | 0.2414348 (12) |
|  | $D\times 10^6$ | 0. 2524(Fix) | 0.2198(17) | 0.2190(Fix) | 0.227(Fix) |
|  | $\gamma$ | -0.002697 (Fix) | -0.000448(1) | -0.002704(Fix) | -0.0004464 (24) |
|  | $b_F$(Yb) | 0.22761(33) | 0.24260(37) | -0.062817(67) | -0.06704 (8) |
|  | $c$(Yb) | 0.0078(14) | 0.009117(12) | -0.00273(45) | -0.002510 (12) |
|  | $e^2Qq_0$(Yb) | N/A | N/A | -0.1107(16) | -0.10996 (6) |
|  | $b_F$(H or F) | 0.000160(fix) | 0.005679(Fix) | 0.000160(fix) | 0.005679(Fix) |
|  | $c$(H or F) | 0.000082(fix) | 0.002849(Fix) | 0.000082(fix) | 0.002849(Fix) |
| $\tilde{A}^2\Pi_{1/2}$ | $A$ | 1350 (Fix) | 1365.3(Fix) | 1350.0(Fix) | 1365. 294 (Fix) |
|  | $B$ | 0.253435(24) | 0.2480568(35) | 0.253185(16) | 0.24779 (6) |
|  | $D\times 10^6$ | 0.2608 (Fix) | 0.2032 (Fix) | 0.2405(Fix) | 0.2032 (Fix) |
|  | $p+2q$ | -0.438667(82) | -0.39762(Fix) | -0.438457(64) | -0.39720 (Fix) |
|  | $a$(Yb) | 0.0148(15) | 0.0128(61) | -0.00422(20) | -0.00507(18) |
|  | $d$(Yb) | 0.03199(58) | 0.0331(16) | -0.00873(13) | -0.00885 (18) |
|  | $e^2Qq_0$(Yb) | N/A | N/A | -0.0642(17) | -0.0647(12) |
|  | $T_{00}$ | 17998.63619(24) | 18788.6502(4) | 17998.60268(13) | 18788.85939 |
|  |  |  |  |  |  |

a) $X^2\Sigma^+(v=0)$ values from Ref.[47]; $A^2\Pi_{1/2}(v=0)$ values from Ref.[56].
b) From combined fit of optical and microwave data (Ref. [46]).

Spectral predictions, such as those presented in Figures 2 – 5, were essential for the assignment and analysis. The approach was identical to that used in modelling the (0,0) $A^2\Pi_{1/2}$- $X\,^2\Sigma^+$ band of $^{171}$YbF(Ref. [47]) and $^{173}$YbF(Ref.[46]). The electric dipole transition moment matrix was constructed in a sequentially coupled Hund's case $(a_{\beta J})$ basis set $|\eta\Lambda\rangle|S\Sigma\rangle|J\Omega(JI(Yb))F_1(F_1I(H))F\rangle$ and cross multiplied by the eigenvectors to produce the transition moments. The relative intensities were taken as the product of the square of the transition moment, a Boltzmann factor and the relative isotope abundance. The relative intensities and a Lorentzian line shape were then used to predict the spectra. The complete set of spectroscopic parameters for the six most abundant isotopologues used in these predictions is presented in Table S3 of the Supplementary Material. The parameter values for $^{171}$YbOH and $^{173}$YbOH were taken from the present study, while those for $^{174}$YbOH($\tilde{X}\,^2\Sigma^+(0,0,0)$) from the analysis of the microwave spectrum[37]. The parameters for $^{174}$YbOH ($\tilde{A}\,^2\Pi_{1/2}(0,0,0)$), $^{172}$YbOH($\tilde{X}\,^2\Sigma^+(0,0,0)$) and $^{172}$YbOH($\tilde{A}\,^2\Pi_{1/2}(0,0,0)$) from the analysis of the optical spectra[35]. The parameters for the $^{170}$YbOH and $^{176}$YbOH isotopologues were obtained by extrapolation of those for $^{171}$YbOH, $^{172}$YbOH, $^{173}$YbOH and $^{174}$YbOH, using the expected mass dependence. In all cases, the $\tilde{X}\,^2\Sigma^+(0,0,0)$ proton magnetic hyperfine parameters were constrained to the values of $^{174}$YbOH, and those for the $\tilde{A}\,^2\Pi_{1/2}(0,0,0)$ constrained to zero.

**DISCUSSION**

The major objectives of this study are to analyze the $0_0^0$ $\tilde{A}\,^2\Pi_{1/2}$ - $\tilde{X}\,^2\Sigma^+$ rotationally resolved spectra of the odd isotopologues, precisely determine the $\tilde{A}\,^2\Pi_{1/2}(0,0,0)$ and $\tilde{X}\,^2\Sigma^+(0,0,0)$ energy levels, and demonstrate the utility of chemical enhancement as a spectroscopic tool. The obtained spectral information is needed for the design and implementation of experiments underway to characterize NSD-PV and search for new T-violating BSM physics. The calculated energies for the $N$=0-4 levels of the $\tilde{X}\,^2\Sigma^+(0,0,0)$ state and $J$=0.5-5.5 levels of the $\tilde{A}\,^2\Pi_{1/2}(0,0,0)$ state for

$^{171}$YbOH are presented in Tables S4 and S5, respectively, of the Supplementary Material. The calculated energies for the $N$=0-4 levels of the $\tilde{X}\,^2\Sigma^+(0,0,0)$ state and $J$=0.5-5.5 levels of the $\tilde{A}\,^2\Pi_{1/2}(0,0,0)$ state for $^{173}$YbOH are presented in Tables S6 and S7, respectively, of the Supplementary Material. Spectroscopic parameters given in Table S3 can be used to accurately predict the energies of higher rotational levels of the even and odd isotopologues. A perusal of the energy level patterns for the odd isotopologues $\tilde{A}\,^2\Pi_{1/2}(0,0,0)$ states reveal that the large Λ-doubling, relative to the rotational spacing, makes the patterns more similar to that of a $^2\Sigma^-$ state than a $^2\Pi_{1/2}$ state. Specifically, the pattern is that of rotationally spaced $J=N\pm1/2$ pairs of levels (i.e. ρ-doublets), with the exception of the $N$=0, $J$=1/2 lowest energy level, which has negative parity. Although the energy levels could be modeled as a $^2\Sigma^-$ state, a $^2\Sigma^+ - {}^2\Sigma^-$ transition is electric dipole forbidden and not consistent with the observed intensities.

The predicted stick spectra for $^{171}$YbOH and $^{173}$YbOH generated using the optimized spectroscopic parameters and a rotational temperature of 15 K are presented in Figures 6 and 7, respectively. Of particular interest are the transitions that terminate to the $J$=0.5, + parity level of the $\tilde{A}\,^2\Pi_{1/2}(0,0,0)$ state, to be used for photon cycling and laser cooling[18]. For the even isotopologues these are the $^PQ_{12}$ (1) and $^PP_{12}$ (1) transitions and for the odd isotopologues they are the $^PQ_{1G}+{}^PP_{1G}$ (1) transitions. These transitions are in a highly congested region of the spectrum (see Figures 6 and 7). It is noteworthy that the $^PQ_{12}$ (1) and $^PP_{12}$ (1) lines of the even isotopologues are "rotationally closed" (i.e. the excitation and fluorescence spectra only involve the $N''$=1 levels) whereas in the odd isotopologues the branching ratio for the ($J$=0.5, + ) $\tilde{A}\,^2\Pi_{1/2}(0,0,0) \to (N=3, -)$ $\tilde{X}\,^2\Sigma^+(0,0,0)$ transition is ~1% relative to the ($J$=0.5, + ) $\tilde{A}\,^2\Pi_{1/2}(0,0,0) \to (N=1,-)\,\tilde{X}\,^2\Sigma^+(0,0,0)$ transitions. This is primarily caused by the mixing of the $J$=0.5, + parity and close lying ($\Delta E \cong 3$

GHz) $J=1.5$, + parity levels of the $\tilde{A}^2\Pi_{1/2}(0,0,0)$ state by hyperfine terms in the effective Hamiltonian. There is a similar, yet much smaller in magnitude, hyperfine induced mixing of the $N=1$ and $N=3$ levels of the $\tilde{X}^2\Sigma^+(0,0,0)$ state.

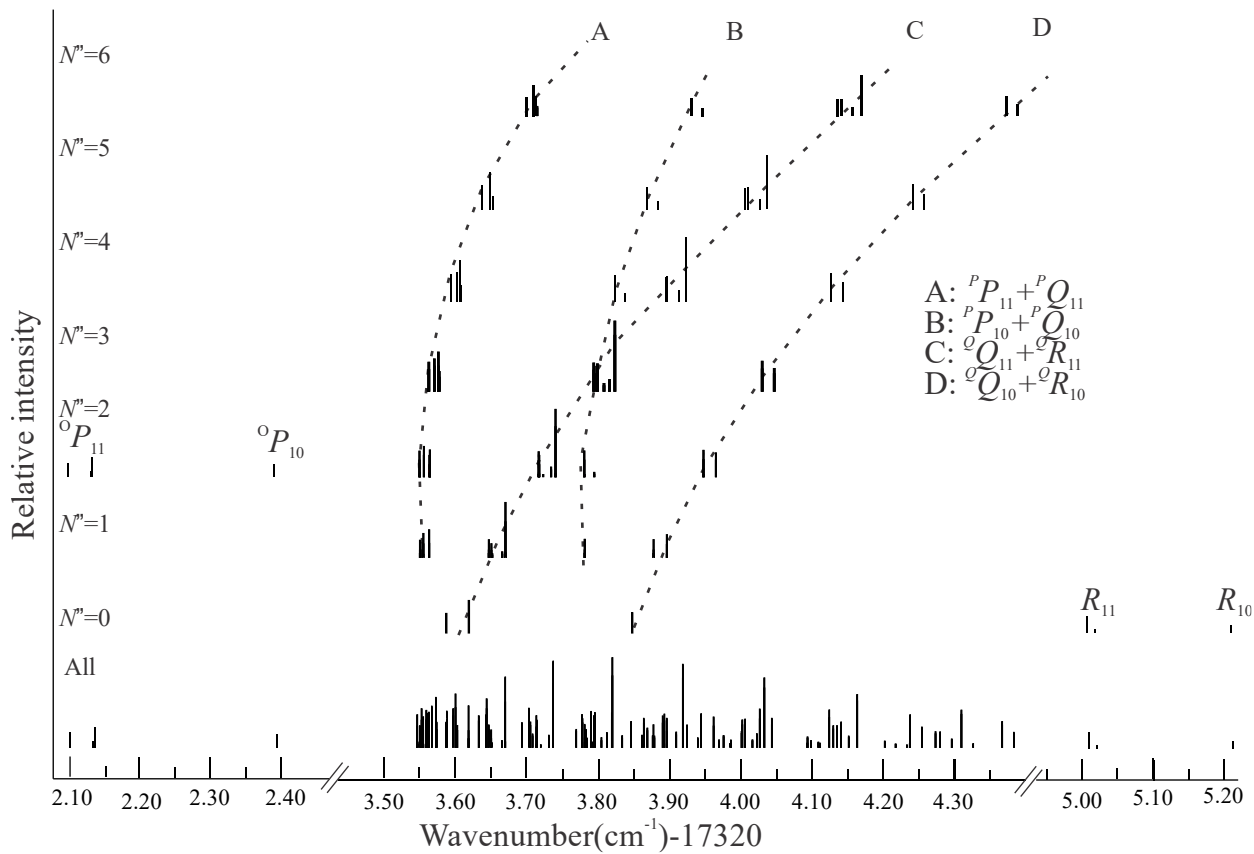

**Figure 6:** The predicted stick spectra with associated assignments for $^{171}$YbOH generated using the optimized spectroscopic parameters of Table I and a rotational temperature of 15 K.

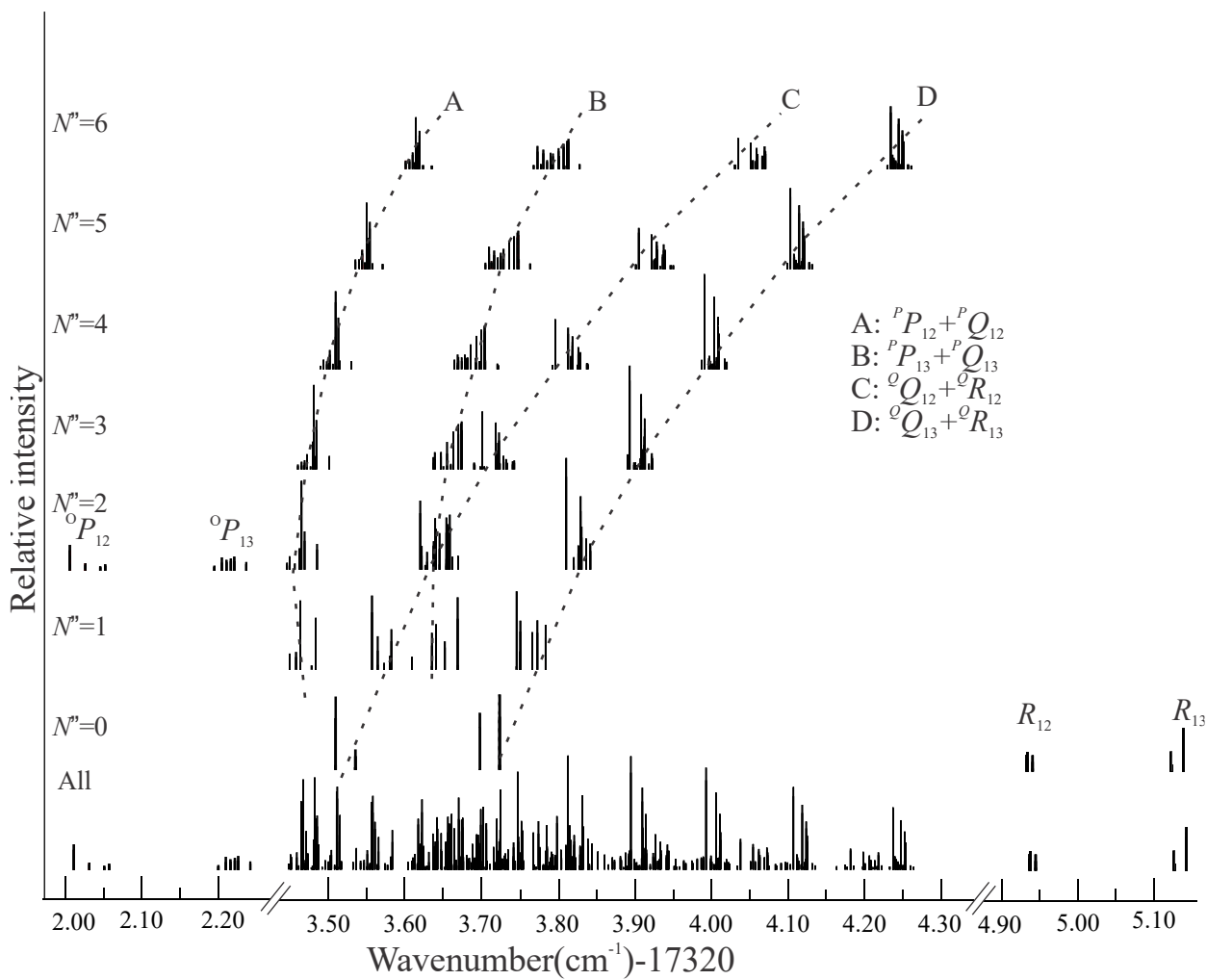

**Figure 7:** The predicted stick spectra with associated assignments for $^{173}$YbOH generated using the optimized spectroscopic parameters of Table I and a rotational temperature of 15 K.

The hyperfine parameters are of particular interest because they are sensitive probes of the electronic wavefunction in the region of the nuclei. The ability to accurately predict hyperfine parameters is the most direct gauge of the of the computational methodology used to predict the $W_d$, $W_Q$, and $W_M$ coupling constants of Eq. 2. As part of the recent calculation[16] of the nuclear magnetic quadrupole interaction constants $W_M$ for YbOH, the ground state $A_{\parallel}$ ($= b_F + \frac{2}{3} c$) magnetic hyperfine constant for $^{173}$YbOH and $^{171}$YbOH, and the axial electric quadrupole coupling constant $e^2Qq_0$ for $^{173}$YbOH were predicted. These predictions were performed using numerical gradients of a four-component Dirac Coulomb Hamiltonian, with electronic correlation treated using the multireference Fock-space coupled cluster method (FSCC). Similarly, a recent calculation[51] of $W_d$ (and the scalar-pseudoscalar nucleon-electron current interactions coupling constant $W_s$) includes predictions for both the ground state $A_\parallel$ and $A_\perp (= b_F - \frac{2}{3}c)$ magnetic hyperfine constants for $^{173}$YbOH. In this case, quasi-relativistic two-component calculations with many-body interactions treated at the level of complex generalized Hartree-Fock (cGHF) and complex generalized Kohn-Sham (cGKS) density functional theory methods were implemented. Most recently, the axial electric quadrupole coupling constant $e^2Qq_0$ for $^{173}$YbOH was predicted using analytic gradients formulated from spin-orbit CCSD(T) theory, computed against an atomic mean-field scalar relativistic (SFX2C-AMF) Hamiltonian[52]. The predicted hyperfine parameters for the $\tilde{X}^2\Sigma^+(0,0,0)$ state are compared with the measured values in Table II. There are no theoretical predictions for the hyperfine interactions in the $\tilde{A}^2\Pi_{1/2}(0,0,0)$ state. The calculated hyperfine parameters obtained using relativistic coupled cluster approachs[16,52] are in excellent agreement with the observed values, with the calculated magnetic hyperfine parameters being 2.5% (Ref. 16) larger in magnitude and the $e^2Qq_0$ parameter being 5.5% (Ref. 16) and 5.2% (Ref. 52) larger in magnitude. Evidently, calculating the core polarization is more difficult than the valence electron properties. Nonetheless, the excellent agreement suggests that the calculated $W_M$ coupling constants for ground state $^{173}$YbOH of Ref. [16], and the effective electric field, $E_{eff}$, of Ref.[53], which was obtained using the same computational methodology, are quantitatively accurate. The predicted $E_{eff}$ is relevant to the electron electric dipole moment (eEDM) measurements for all the isotopologues of YbOH. The quasirelativistic two component calculations in Ref. 51 give magnetic hyperfine constants that are smaller in magnitude by approximately 15% and 31% for the cGHF and cGKS methods, respectively.

**Table II**: Comparison of the measured hyperfine parameters of the $\tilde{X}^2\Sigma^+(0,0,0)$ state of $^{171,173}$YbOH to calculated values

| Isotopologue | Parameter | Measured (MHz) | Theory Ref. [16] (MHz) | Theory Ref. [51] cGHF(MHz) | Theory Ref. [51] cGKS(MHz) | Theory Ref.[52] (MHz) |
|---|---|---|---|---|---|---|
| $^{171}$YbOH | $A_\parallel$ [a] | 6979 (35) | 7174.9 | | | |
| $^{171}$YbOH | $A_\perp$ [b] | 6745(15) | | | | |
| $^{173}$YbOH | $A_\parallel$ | -1929(11) | −1976.3 | -1600 | -1300 | |
| $^{173}$YbOH | $A_\perp$ | -1856 (5) | | -1600 | | |
| $^{173}$YbOH | $e^2Qq_0$ | -3319 (48) | −3502 | | | -3492 |

a: For a $\sigma$ orbital: $A_\parallel = b + c = b_F + \frac{2}{3}c$

b: For a $\sigma$ orbital: $A_\perp = b = b_F - \frac{1}{3}c$

Although the high-level relativistic electronic structure calculation[16] accurately predicts the hyperfine interactions for the $\tilde{X}^2\Sigma^+(0,0,0)$ state, a less quantitative, molecular-orbital based, model using atomic information is also useful. Such a model provides chemical insight, can be used to rationalize trends amongst similar molecules (e.g. YbF and YbOH), and is able to predict hyperfine interactions for excited electronic states that are not readily addressed by relativistic electronic structure calculations. The $\tilde{X}^2\Sigma^+(0,0,0)$ hyperfine fitting parameters are related to various averages over the spatial coordinates of the electrons by[49,54]:

$$b_F / Hz = \left(\frac{\mu_0}{4\pi h}\right)\left(\frac{8\pi}{3}\right) g_e g_N \mu_B \mu_N \frac{1}{S} \langle \Lambda\Sigma = S | \sum_i \hat{s}_{zi} \delta_i(r) | \Lambda\Sigma = S \rangle \qquad (3)$$

$$c / Hz = \left(\frac{\mu_0}{4\pi h}\right)\left(\frac{3}{2}\right) g_e g_N \mu_B \mu_N \frac{1}{S} \langle \Lambda\Sigma = S | \sum_i \hat{s}_{zi} \frac{(3\cos^2\theta_i - 1)}{r_i^3} | \Lambda\Sigma = S \rangle \qquad (4)$$

and

$$e^2 Q q_0 / Hz = -Q \frac{e^2}{4\pi\varepsilon_0 h} \langle \Lambda | \sum_i \frac{(3\cos^2\theta_i - 1)}{r_i^3} | \Lambda \rangle, \quad (5)$$

where $\hat{s}_{zi}$ is the spin angular momentum operators for the $i^{th}$ electron, $\delta_i(r)$ is a Dirac delta function, and $r$ and $\theta$ are polar coordinates. In the case of $b_F$ and $c$, the sum runs only over unpaired electrons, whereas for $e^2 Q q_0$ the sum is over all electrons. The observed ratio $b_F(^{171}\text{YbOH})/b_F(^{173}\text{YbOH})$ for $\tilde{X}\,^2\Sigma^+(0,0,0)$ state ($= -3.62 \pm 0.04$) is in excellent agreement with that of the nuclear g-factors, $g_N(^{171}\text{Yb})/g_N(^{173}\text{Yb})$ ($= -3.630$). Meanwhile the ratio for the less well determined dipolar ratio $c(^{171}\text{YbOH})/c(^{173}\text{YbOH})$ ($= -2.86 \pm 0.57$) is within two standard deviations. To a first approximation, the dominant electronic configuration for the $\tilde{X}\,^2\Sigma^+(0,0,0)$ state has one unpaired electron in a hybridized $6s/6p/5d$ σ-type, Yb$^+$- centered orbital which is polarized away from the Yb-OH bond. The hybridization is driven by the stabilization achieved from shifting the center of charge for the unpaired electron away from the electrophilic end of the Yb$^+$OH$^−$ molecule. Based upon a comparison with $b_F(^{171}\text{Yb}^+)$ ($= 0.4217$ cm$^{-1}$)[55] for Yb$^+$($f^{14} 6s\ ^2S_{1/2}$), the σ-type orbital is approximately 54% $6s$ character. The $b_F(^{171}\text{YbOH})$ and $b_F(^{173}\text{YbOH})$ values are approximately 7 % smaller in magnitude than the corresponding values for YbF. Evidently, the unpaired $6s$ electron is more effectively polarized by OH$^-$ than F$^-$ (i.e. larger $6s/6p/5d$ hybridization). This is consistent with the point charge electrostatic model prediction that successfully models the permanent electric dipole moments for ground states of YbOH and YbF[35]. The Yb$^+$-centered, unpaired electron was predicted to be shifted away from the Yb-F and Yb-O bond by 0.859 and 0.969 Å, respectively, reflecting the increased $6s/6p/5d$ hybridization in YbOH relative to YbF.

Unlike the noted change in $b_F$, the $^{173}$Yb axial nuclear electric quadrupole parameter, $e^2 q_0 Q$, for the ground states of $^{173}$YbOH and $^{173}$YbF are very nearly identical, indicating that core polarization makes a substantial contribution relative to that of the unpaired valence electron. As evident from a comparison of Eqs. 4 and 5, the contribution from the unpaired electron to $e^2q_0Q$ is:

$$e^2Qq_0(unpaired) = -c\left(\frac{Qe^22}{3g_Ng_e\mu_0\mu_N}\right). \quad (6)$$

Substitution of $Q=280$ fm$^2$ and $g_N$ of -0.27195 gives $e^2Qq_0(unpaired) = -0.0456$ cm$^{-1}$, which is approximately 40% (-0.0456/-0.1107) of the determined $e^2Qq_0$ value for $\tilde{X}\,^2\Sigma^+(0,0,0)$ state. The remaining -0.0651 cm$^{-1}$ portion is due to core polarization.

An interpretation of the excited state hyperfine parameters is more difficult because $b_F$ and $c$ were constrained to zero. The Λ-doubling type magnetic hyperfine parameter, $d$, on the other hand, is well determined and is given by[49,54]:

$$d/Hz = \left(\frac{\mu_0}{4\pi h}\right)\left(\frac{3}{2}\right)g_e g_N \mu_B \mu_N \frac{1}{\sqrt{S(S+1)-\Sigma(\Sigma+1)}}\langle\Lambda=-1;S\Sigma+1|\sum_i \hat{s}_i^+ \frac{(\sin^2\theta_i e^{-2i\phi_i})}{r_i^3}|\Lambda=+1;S\Sigma\rangle, \quad (7)$$

where $\hat{s}_i^+$ is the one electron spin angular momentum raising operator and $\phi_i$ is the azimuthal angle of the electron. The ratio of $d$ for $^{171}$YbOH and $^{173}$YbOH (=0.03199/-0.00873 ≅ 3.66) is in excellent agreement with expected ratio of $g_N$ values. The same rationale presented for the observed $d$ values of YbF[56] is applicable to YbOH, and an estimate gives values for $^{171}$YbOH and $^{173}$YbOH of 0.0657 and -0.0182 cm$^{-1}$, which are the correct sign but approximately a factor of two too large. These values were obtained by assuming that the orbital of the sole unpaired electron for the $\tilde{A}\,^2\Pi_{1/2}(0,0,0)$ state is pure Yb$^+$ 6$p$. The more realistic assumption that the orbital of the sole unpaired electron is an admixture of Yb$^+$ 6p, Yb 6p, and Yb 5d and would give values in closer agreement with the experimental values.

The magnitude of the $e^2q_0Q$ parameter decreases from 0.1107(16) cm$^{-1}$ to 0.0642(17) cm$^{-1}$ upon $\tilde{X}^2\Sigma^+(0,0,0)$ to $\tilde{A}^2\Pi_{1/2}(0,0,0)$ excitation. The angular expectation value for the 6p orbital contribution to the open shell σ-orbital of the $\tilde{X}^2\Sigma^+(0,0,0)$ state is $\langle p_0|(3\cos^2\theta_i - 1)|p_0\rangle = 4/5$ while that for the open shell π-orbital of the $\tilde{A}^2\Pi_{1/2}(0,0,0)$ state is $\langle p_{\pm 1}|(3\cos^2\theta_i - 1)|p_{\pm 1}\rangle = -2/5$. Accordingly, if the σ- and π-orbitals of the two states had identical 6p contribution, then $e^2Qq_0(unpaired)$ for $\tilde{A}^2\Pi_{1/2}(0,0,0)$ would be approximately 0.0228 cm$^{-1}$ (i.e. half the magnitude and opposite sign of the $\tilde{X}^2\Sigma^+(0,0,0)$ state value). Assuming that the core polarization contributions in the $\tilde{A}^2\Pi_{1/2}(0,0,0)$ state is the same as the $\tilde{X}^2\Sigma^+(0,0,0)$ state, then $e^2Qq_0$ for the $\tilde{A}^2\Pi_{1/2}(0,0,0)$ state is predicted to be -0.0423 cm$^{-1}$, in qualitative agreement with observed values given the assumptions being made.

**CONCLUSION**

A novel spectroscopic technique utilizing laser-induced chemical reactions to distinguish features in congested and overlapped spectra was demonstrated. This technique was utilized to determine the fine and hyperfine structure of $^{171,173}$YbOH for the first time. The derived molecular parameters are consistent with previous measurements of the even isotopologues $^{172,174}$YbOH as well as with isoelectronic $^{171,173}$YbF. The determined hyperfine parameters also provide an experimental comparison to check the quality of the calculated P- and T,P-violating coupling constants $W_P$, $W_d$, and $W_M$. Finally, the characterization of the $\tilde{X}^2\Sigma^+(0,0,0) - \tilde{A}^2\Pi_{1/2}(0,0,0)$ transition of $^{171,173}$YbOH performed in this work will aid in the implementation of NSD-PV and NMQM measurements, laser cooling and photon cycling, and future spectroscopic investigations of these promising molecular isotopologues.

## SUPPLEMENTARY MATERIAL

See supplementary material for the Description: Description of the effective Hamiltonian and Seven tables. Table S1: The transition wavenumber for the $A\,^2\Pi_{1/2} - X\,^2\Sigma^+$ (0,0,0) electronic transition of $^{171}$YbOH. Table S2: The transition wavenumber for the $A\,^2\Pi_{1/2} - X\,^2\Sigma^+$ (0,0,0) electronic transition of $^{173}$YbOH. Table S3: Spectroscopic parameters in wavenumbers (cm$^{-1}$) used to model the $\tilde{X}\,^2\Sigma^+(0,0,0)$ and $\tilde{A}\,^2\Pi_{1/2}(0,0,0)$ states of YbOH. Table S4: The calculated low-rotational ($N$=0-4) $\tilde{X}\,^2\Sigma^+(0,0,0)$ state energies of $^{171}$YbOH. Table S5: The calculated low-rotational ($J$=0.5-5.5) $\tilde{A}\,^2\Pi_{1/2}(0,0,0)$ state energies of $^{171}$YbOH. Table S6: The calculated low-rotational ($N$=0-4) $\tilde{X}\,^2\Sigma^+(0,0,0)$ state energies of $^{173}$YbOH. Table S7: The calculated low-rotational ($J$=0.5-5.5) $\tilde{A}\,^2\Pi_{1/2}(0,0,0)$ state energies of $^{173}$YbOH.

## ACKNOWLEDGMENT

The research at Arizona State University and Caltech was supported by grants from the Heising-Simons Foundation (ASU: Grant 2018-0681; Caltech: Grant 2019-1193). NRH acknowledges support from a NIST Precision Measurement Grant (60NANB18D253) and NSF CAREER Award (PHY-1847550). We thank Graceson Aufderheide and Richard Mawhorter (Physics and Astronomy Department, Pomona College, Pomona, CA 91711, USA) for assistance with Yb+Te(OH)$_6$ target fabrication, Phelan Yu for help fabricating the Yb+Yb(OH)$_3$ targets, and Dr. Anh Le (Chemistry Dept. Georgia Institute of Technology , Atlanta GA 30318, USA) for assistance in recording the molecular beam spectra. We thank Ben Augenbraun and Phelan Yu for feedback on the manuscript.

**SUPPLEMENTARY MATERIAL**

**The Effective Hamiltonian Model**

The effective Hamiltonian models used for the $\tilde{X}\,^2\Sigma^+(0,0,0)$ and $\tilde{A}\,^2\Pi_{1/2}(0,0,0)$ state were taken as:

$$\hat{H}^{\text{eff}}(\tilde{X}\,^2\Sigma^+) = B\hat{R}^2 - D\hat{R}^2\hat{R}^2 + \gamma\hat{N}\cdot\hat{S}$$
$$+ b_F(^{171,173}\text{Yb,H})\hat{I}\cdot\hat{S} + c(^{171,173}\text{Yb,H})\frac{1}{3}\left(3\hat{I}_z\hat{S}_z - \hat{I}\cdot\hat{S}\right) + e^2Qq_0(^{173}\text{Yb})\frac{(3\hat{I}_z^2 - \hat{I}^2)}{4I(2I-1)} \quad (1)$$

and

$$\hat{H}^{\text{eff}}(\tilde{A}\,^2\Pi_{1/2}) = A\hat{L}_z\hat{S}_z + B\hat{R}^2 - D\hat{R}^2\hat{R}^2 - \frac{1}{2}(p+2q)(\hat{J}_-\hat{S}_-e^{+2i\phi} + \hat{J}_+\hat{S}_+e^{-2i\phi})$$
$$+ a(^{171,173}\text{Yb})\hat{I}_z\hat{L}_z - \frac{1}{2}d(^{173}\text{Yb},^{19}\text{F})(\hat{S}_+\hat{I}_+e^{-2i\phi} + \hat{S}_-\hat{I}_-e^{+2i\phi}) + e^2Qq_0(^{173}\text{Yb})\frac{(3\hat{I}_z^2 - \hat{I}^2)}{4I(2I-1)}, \quad (2)$$

where $\hat{O}_\pm$ are shift operators and the coordinate $\phi$ is the electronic azimuthal angle associated with the orbital angular momentum about the $z$-axis. The exponential term involving $\phi$ assures that $\Lambda$-doubling ($p+2q$) term and the hyperfine $d$ term connect states whose $\Lambda$ values differ by $\pm 2$. In the analysis performed here the standard spherical harmonic phase choice has been assumed:

$$\langle\Lambda = \pm 1|e^{\pm 2i\phi}|\Lambda = \mp 1\rangle = -1. \quad (3)$$

The Fermi contact, $b_F$, and dipolar, $c$ terms for the $\tilde{A}\,^2\Pi_{1/2}(0,0,0)$ state were omitted because the data set, which is restricted to $\tilde{A}\,^2\Pi_{1/2}(0,0,0)$ spin-orbit component, is only sensitive to the combination $a - \frac{b_F}{2} - \frac{c}{3}$ ($\equiv h_{1/2}$).

Eqs. 1 and 2 in spherical tensor form are:

$$\hat{H}^{eff}(\tilde{X}\,^2\Sigma^+) = BT^1(\hat{R})\cdot T^1(\hat{R}) - D\left[T^1(\hat{R})\cdot T^1(\hat{R}), T^1(\hat{R})\cdot T^1(\hat{R})\right]_+ + \gamma\left\{T^1(\hat{J})\text{-}T^1(\hat{S})\right\}\cdot T^1(\hat{S})$$

$$+ b_F(^{171,173}\text{Yb},\text{H})T^1(\hat{I})\cdot T^1(\hat{S}) + c(^{171,173}\text{Yb},\text{H})\frac{1}{3}\sqrt{6}T^2_{q=0}(\hat{I},\hat{S}) \quad (4)$$

$$+ e^2Qq_0(^{173}\text{Yb})\sqrt{\frac{3}{2}}\frac{1}{2I(2I-1)}T^2_{q=0}(\hat{I},\hat{I})$$

and

$$\hat{H}^{eff}(\tilde{A}\,^2\Pi_{1/2}) = AT^1_0(\hat{L})T^1_0(\hat{S}) + BT^1(\hat{R})\cdot T^1(\hat{R}) - D\left[T^1(\hat{R})\cdot T^1(\hat{R}), T^1(\hat{R})\cdot T^1(\hat{R})\right]_+ - (p+2q)\sum_{q=\pm1}e^{-2iq\phi}T^2_{2q}(\hat{J},\hat{S})$$

$$+ a(^{171,173}\text{Yb})T^1_0(\hat{I})T^1_0(\hat{L}) - d(^{173}\text{Yb},^{19}\text{F})\sum_{q=\pm1}e^{-2iq\phi}T^2_{2q}(\hat{I},\hat{S})$$

$$+ e^2Qq_0(^{173}\text{Yb})\sqrt{\frac{3}{2}}\frac{1}{2I(2I-1)}T^2_{q=0}(\hat{I},\hat{I})$$

$$(5)$$

where $[\ ]_+$ is the anticommutator. The orbital hyperfine operator is more completely written as $aT^1(\hat{I})\cdot T^1(\hat{L})$ which has matrix elements between Hund's case (a) electronic states (e.g. the $\tilde{A}\,^2\Pi_{1/2}(0,0,0)$ and $\tilde{B}\,^2\Sigma^+(0,0,0)$ states) and are ignored in the present analysis. Similiarly, only the $c$ and $d$ terms of the dipolar interaction are used in the model. The dipolar magnetic hyperfine Hamiltonian is:

$$\hat{H}^{eff}(dipolar) = -\sqrt{10}g_S\mu_B g_N(\mu_0/4\pi)T^1(\hat{S},\hat{C}^2)\cdot T^1(\hat{I})$$
$$= \sqrt{6}g_S\mu_B g_N(\mu_0/4\pi)\sum_{q_2}(-1)^{q_2}T^2_{q_2}(\hat{C}^2)T^2_{-q_2}(\hat{I},\hat{S}), \quad (6)$$

where $T^2_{q_2}(\hat{C}^2)$ are the modified spherical harmonics ($=\sqrt{\frac{4\pi}{5}}Y^2_l(\theta,\varphi)$). The $c$ and $d$ terms are associated with $q_2=0$ and $\pm2$ components of $T^2_{q_2}(\hat{C}^2)$. The ignored $T^2_{\pm1}(\hat{C}^2)$ term mixes Hund's case (a) electronic states that having $\Delta\Lambda=\pm1$. Similarly, there are $q=\pm1$ and $\pm2$ related terms for the nuclear electric quadrupole interacrtions which have been ignored.

The angular momenta that appear in Eqs. 1, 2, 3 and 4 are:

$\vec{R}=\vec{J}-\vec{L}-\vec{S}$ : Rigid nuclear framework rotation,

$\vec{L}$ and $\vec{S}$ : Total electronic orbital and electron spin,

$\vec{N}=\vec{R}+\vec{L}$ : Total orbital,

$\vec{J}=\vec{R}+\vec{L}+\vec{S}$ : Total in the absence of nuclear spin,

and

$\vec{I}$ : Nuclear spin.

**Table S1.** The transition wavenumbers and assignments for the $\tilde{A}^2\Pi_{1/2}(0,0,0) - \tilde{X}^2\Sigma^+(0,0,0)$ band of $^{173}$YbOH.

| Line | $N'', G'', F'', p$ | $J', F', p$ | Obs. (cm$^{-1}$) | Obs.-calc. (MHz) |
|---|---|---|---|---|
| $^OP_{12}$ | 2, 2, 4, + | 0.5, 3, − | 17322.0363 | -13 |
|  | 2, 2, 2, + | 0.5, 3, − | 22.0504 | -1 |
|  | 2, 2, 3, + | 0.5, 3, − | 22.0566 | -8 |
|  | 2, 2, 2, + | 0.5, 2, − | 22.0771 | 29 |
|  | 2, 2, 3, + | 0.5, 2, − | 22.0826 | 0 |
|  | 3, 2, 5, − | 1.5, 4, + | 21.1070 | 5 |
|  | 3, 2, 4, − | 1.5, 4, + | 21.1258 | -33 |
|  | 3, 2, 4, − | 1.5, 3, + | 21.1303 | -45 |
|  | 3, 2, 3, − | 1.5, 2, + | 21.1408 | -24 |
|  | 3, 2, 2, − | 1.5, 1, + | 21.1460 | -28 |
|  | 4, 2, 6, + | 2.5, 5, − | 20.1905 | -24 |
|  | 4, 2, 5, + | 2.5, 4, − | 20.2117 | 22 |
|  | 4, 2, 4, + | 2.5, 3, − | 20.2202 | 12 |
| $^OP_{13}$ | 2, 3, 2, + | 0.5, 3, − | 22.2251 | -6 |
| | 2, 3, 3, + | 0.5, 3, − | 22.2407 | -7 |
| | 2, 3, 4, + | 0.5, 3, − | 22.2463 | -14 |
| | 2, 3, 3, + | 0.5, 2, − | 22.2664 | -8 |
| | 3, 3, 1, − | 1.5, 2, + | 21.2980 | -3 |
| | 3, 3, 2, − | 1.5, 2, + | 21.3090 | -10 |
| | 3, 3, 5, − | 1.5, 4, + | 21.3176 | 17 |
| | 3, 3, 4, − | 1.5, 3, + | 21.3223 | 32 |
| | 4, 3, 2, + | 2.5, 2, − | 20.3900 | -15 |
| | 4, 3, 3, + | 2.5, 3, − | 20.3900 | -33 |
| | 4, 3, 4, + | 2.5, 4, − | 20.3951 | -9 |
| | 4, 3, 4, + | 2.5, 3, − | 20.4020 | -16 |
| | 4, 3, 3, + | 2.5, 2, − | 20.4020 | -9 |
| | 4, 3, 5, + | 2.5, 4, − | 20.4020 | -28 |
| $^PQ_{12}+^PP_{12}$ | 1, 2, 1, − | 0.5, 2, + | 23.4572 | 37 |
| | 1, 2, 3, − | 0.5, 2, + | 23.4654 | 35 |
| | 1, 2, 2, − | 0.5, 3, + | 23.4916 | 45 |
| | 2, 2, 1, + | 1.5, 2, − | 23.4572 | 9 |
| | 2, 2, 4, + | 1.5, 4, − | 23.4725 | 0 |
| | 2, 2, 3, + | 1.5, 3, − | 23.4783 | 58 |
| | 3, 2, 5, − | 2.5, 5, + | 23.4877 | 13 |
| | 3, 2, 4, − | 2.5, 4, + | 23.4916 | 34 |
| | 4, 2, 4, + | 3.5, 4, − | 23.5181 | 42 |
| | 4, 2, 6, + | 3.5, 6, − | 23.5181 | 46 |
| | 4, 2, 5, + | 3.5, 5, − | 23.5212 | 28 |
| | 5, 2, 7, − | 4.5, 7, + | 23.5642 | 80 |
| | 5, 2, 5, − | 4.5, 5, + | 23.5642 | 46 |
| | 6, 2, 7, + | 5.5, 7, − | 23.6277 | 35 |
| $^PQ_{13}+^PP_{13}$ | 1, 3, 2, − | 0.5, 3, + | 23.6474 | 1 |
| | 1, 3, 3, − | 0.5, 2, + | 23.6762 | 21 |
| | 2, 3, 2, + | 1.5, 1, − | 23.6535 | 29 |
| | 2, 3, 3, + | 1.5, 2, − | 23.6633 | 56 |
| | 2, 3, 4, + | 1.5, 2, − | 23.6672 | 28 |
| | 3, 3, 2, − | 1.5, 3, + | 23.6535 | -3 |
| | 3, 3, 3, − | 2.5, 3, + | 23.6701 | 21 |
| | 3, 3, 4, − | 2.5, 2, + | 23.6762 | -4 |
| | 3, 3, 6, − | 2.5, 3, + | 23.6802 | 10 |
| | 3, 3, 5, − | 2.5, 4, + | 23.6802 | -11 |
| | 4, 3, 5, + | 3.5, 4, − | 23.7070 | 5 |
| | 4, 3, 6, + | 3.5, 5, − | 23.7114 | 9 |
| | 4, 3, 7, + | 3.5, 6, − | 23.7114 | 1 |
| | 5, 3, 4, − | 4.5, 3, + | 23.7396 | 8 |
| | 5, 3, 5, − | 4.5, 4, + | 23.7465 | -4 |
| | 5, 3, 6, − | 4.5, 5, + | 23.7534 | 12 |
| | 5, 3, 7, − | 4.5, 6, + | 23.7580 | 17 |
| | 5, 3, 8, − | 4.5, 7, + | 23.7580 | -11 |
| $^QR_{12}+^QQ_{12}$ | 0, 2, 2, + | 0.5, 3, − | 23.5181 | 53 |
| | 1, 2, 3, − | 1.5, 4, + | 23.5642 | 9 |
| | 1, 2, 2, − | 1.5, 3, + | 23.5879 | -35 |
| | 1, 2, 1, − | 1.5, 1, + | 23.5879 | 18 |
| | 2, 2, 4, + | 2.5, 5, − | 23.6277 | -10 |
| | 2, 2, 3, + | 2.5, 4, − | 23.6474 | -5 |
| | 2, 2, 2, + | 2.5, 3, − | 23.6474 | -26 |
| | 3, 2, 5, − | 3.5, 6, + | 23.7070 | -36 |
| | 3, 2, 4, − | 3.5, 5, + | 23.7256 | -6 |
| | 3, 2, 1, − | 3.5, 1, + | 23.7351 | -22 |
| | 4, 2, 6, + | 4.5, 7, | 23.8020 | -70 |
| | 4, 2, 2, + | 4.5, 2, | 23.8362 | -38 |
| | 5, 2, 7, − | 5.5, 8, + | 23.9167 | 11 |
| | 5, 2, 6, − | 5.5, 7, + | 23.9315 | -40 |
| | 5, 2, 5, − | 5.5, 6, + | 23.9387 | -24 |
| | 5, 2, 4, − | 5.5, 5, + | 23.9387 | -44 |
| | 5, 2, 3, − | 5.5, 3, + | 23.9480 | -16 |
| | 10, 2, 9, + | 10.5, 9, − | 24.7546 | 20 |
| | 10, 2, 8, + | 10.5, 8, − | 24.7546 | -26 |
| $^QR_{13}+{}^QQ_{13}$ | 0, 3, 3, + | 0.5, 2, − | 23.7305 | 1 |
| | 1, 3, 4, − | 1.5, 4, + | 23.7534 | 14 |
| | 1, 3, 4, − | 1.5, 3, + | 23.7580 | 6 |
| | 1, 3, 2, − | 1.5, 1, + | 23.7734 | 9 |
| | 1, 3, 3, − | 1.5, 3, + | 23.7802 | 4 |
| | 1, 3, 3, − | 1.5, 2, + | 23.7907 | -6 |
| | 2, 3, 5, + | 2.5, 5, − | 23.8189 | 7 |
| | 2, 3, 4, + | 2.5, 4, − | 23.8362 | 53 |
| | 2, 3, 4, + | 2.5, 3, − | 23.8440 | -18 |
| | 2, 3, 3, + | 2.5, 2, − | 23.8502 | 7 |
| | 3, 3, 6, − | 3.5, 6, + | 23.9012 | 12 |
| | 3, 3, 5, − | 3.5, 5, + | 23.9167 | 23 |
| | 3, 3, 4, − | 3.5, 4, + | 23.9219 | 44 |
| | 4, 3, 7, + | 4.5, 7, − | 23.9985 | -14 |
| | 4, 3, 2, + | 4.5, 3, − | 24.0052 | 4 |
| | 4, 3, 1, + | 4.5, 2, − | 24.0052 | -4 |
| | 4, 3, 6, + | 4.5, 6, − | 24.0111 | -26 |
| | 4, 3, 5, + | 4.5, 5, − | 24.0168 | -10 |
| | 4, 3, 4, + | 4.5, 4, − | 24.0168 | -38 |
|  | 5, 3, 8, − | 5.5, 8, + | 24.1125 | -20 |
|  | 5, 3, 2, − | 5.5, 3, + | 24.1192 | 47 |
|  | 5, 3, 7, − | 5.5, 7, + | 24.1254 | 21 |
|  | 5, 3, 6, − | 5.5, 6, + | 24.1305 | 20 |
|  | 5, 3, 5, − | 5.5, 5, + | 24.1305 | -22 |
|  | 9, 3, 11, − | 9.5, 11, + | 24.7284 | -3 |
|  | 9, 3, 10, − | 9.5, 10, + | 24.7411 | 19 |
|  | 9, 3, 9, − | 9.5, 9, + | 24.7411 | -11 |
|  | 9, 3, 8, − | 9.5, 8, + | 24.7411 | 11 |
| $^{R}R_{12}$ | 0, 2, 2, + | 1.5, 3, | 24.9356 | -7 |
|  | 0, 2, 2, + | 1.5, 2, | 24.9356 | -36 |
|  | 0, 2, 2, + | 1.5, 1, | 24.9432 | -11 |
|  | 1, 2, 1, − | 2.5, 2, + | 25.9187 | 3 |
|  | 1, 2, 1, − | 2.5, 1, + | 25.9239 | 19 |
|  | 1, 2, 3, − | 2.5, 3, + | 25.9239 | 1 |
|  | 1, 2, 1, − | 2.5, 0, + | 25.9272 | 24 |
|  | 1, 2, 3, − | 2.5, 4, + | 25.9272 | -7 |
|  | 1, 2, 2, − | 2.5, 3, + | 25.9448 | 18 |
|  | 2, 2, 0, + | 3.5, 1, − | 26.9300 | -7 |
|  | 2, 2, 4, + | 3.5, 5, − | 26.9355 | -36 |
|  | 2, 2, 1, + | 3.5, 2, − | 26.9355 | -18 |
|  | 2, 2, 1, + | 3.5, 1, − | 26.9401 | -5 |
|  | 2, 2, 2, + | 3.5, 3, − | 26.9456 | -16 |
|  | 2, 2, 3, + | 3.5, 4, − | 26.9509 | -17 |
| $^{R}R_{13}$ | 0, 3, 3, + | 1.5, 3, − | 25.1245 | 8 |
|  | 0, 3, 3, + | 1.5, 4, − | 25.1398 | -21 |
|  | 1, 3, 4, − | 2.5, 4, + | 26.1179 | 43 |
|  | 1, 3, 4, − | 2.5, 5, + | 26.1331 | -8 |

|  | 1, 3, 3, − | 2.5, 4, + | 26.1383 | −13 |
|  | 2, 3, 3, − | 3.5, 4, + | 27.1351 | −13 |
|  | 2, 3, 5, − | 3.5, 6, + | 27.1440 | −1 |
| Std. dev. of fit: 25 MHz (0.00082 cm$^{-1}$) (**128 lines to 94 features**) | | | | |
|  |  |  |  |  |

**Table S2.** The transition wavenumber and assignments for the $A\,^2\Pi_{1/2} - X\,^2\Sigma^+$ (0,0) band of $^{171}$YbOH.

| Line | $N'', G'', F'', p$ | $J', F', p$ | Obs. (cm$^{-1}$) | Obs.-calc. (MHz) |
|---|---|---|---|---|
| $^OP_{11}$ | 2, 1, 1, + | 0.5, 0, − | 17322.1144 | 1 |
|  | 2, 1, 2, + | 0.5, 1, − | 22.1467 | 35 |
|  | 3, 1, 2, − | 1.5, 1, + | 21.1929 | −5 |
|  | 3, 1, 3, − | 1.5, 2, + | 21.2133 | −4 |
|  | 4, 1, 3, + | 2.5, 2, − | 20.2756 | −27 |
|  | 4, 1, 4, + | 2.5, 3, − | 20.2969 | −6 |
| $^OP_{10}$ | 2, 0, 2, + | 0.5, 1, − | 22.3760 | 5 |
|  | 3, 0, 3, − | 1.5, 2, + | 21.4437 | −4 |
|  | 4, 0, 4, + | 2.5, 3, − | 20.5263 | −39 |
| $^PP_{11} + {}^PQ_{11}$ | 1, 0, 2 − | 0.5, 1, + | 23.5567 | −25 |
|  | 2, 1, 2, + | 1.5, 2, − | 23.5500 | 12 |
|  | 2, 1, 3, + | 1.5, 2, − | 23.5567 | 42 |
|  | 2, 1, 1, + | 1.5, 1, − | 23.5641 | 1 |
|  | 3, 1, 3, − | 2.5, 3, + | 23.5641 | 53 |
|  | 3, 1, 4, − | 2.5, 3, + | 23.5706 | 32 |
|  | 3, 1, 2, − | 2.5, 2, + | 23.5769 | 40 |
|  | 4, 1, 4, + | 3.5, 4, − | 23.5905 | −18 |
|  | 4, 1, 5, + | 3.5, 4, − | 23.5997 | 0 |
|  | 4, 1, 3, + | 3.5, 3, − | 23.6034 | 7 |
|  | 4, 1, 4, + | 3.5, 3, − | 23.6045 | -50 |
|  | 5, 1, 5, − | 4.5, 5, + | 23.6357 | 0 |
|  | 5, 1, 4, − | 4.5, 4, + | 23.6470 | 12 |
|  | 6, 1, 6, + | 5.5, 6, − | 23.6966 | 16 |
|  | 6, 1, 5, + | 5.5, 5, − | 23.7056 | -7 |
|  | 6, 1, 7, + | 5.5, 6, − | 23.7088 | 33 |
|  | 7, 1, 7, − | 6.5, 7, + | 23.7715 | -27 |
|  | 7, 1, 6, − | 6.5, 6, + | 23.7802 | -22 |
| $^PP_{10}+\,^PQ_{10}$ | 1, 0, 1, − | 0.5, 1, + | 23.7836 | -6 |
|  | 2, 0, 2, + | 1.5, 2, − | 23.7802 | 2 |
|  | 3, 0, 3, − | 2.5, 3, + | 23.7836 | 9 |
|  | 4, 0, 4, + | 3.5, 4, − | 23.8212 | -12 |
|  | 4, 0, 4, + | 3.5, 3, − | 23.8354 | -39 |
|  | 5, 0, 5, − | 4.5, 5, + | 23.8665 | 4 |
|  | 5, 0, 5, − | 4.5, 4, + | 23.8824 | 24 |
|  | 6, 0, 6, + | 5.5, 6, − | 23.9280 | 30 |
|  | 6, 0, 6, + | 5.5, 5, − | 23.9424 | 2 |
| $^QQ_{11}+\,^QR_{11}$ | 0, 1, 1, + | 0.5, 1, − | 23.6220 | 25 |
|  | 1, 1, 2, − | 1.5, 2, + | 23.6738 | 34 |
|  | 2, 1, 2, + | 2.5, 2, − | 23.7164 | -1 |
|  | 2, 1, 2, + | 2.5, 3, − | 23.7329 | -33 |
|  | 2, 1, 3, + | 2.5, 3, − | 23.7384 | -40 |
|  | 3, 1, 3, − | 3.5, 3 + | 23.7969 | -41 |
|  | 3, 1, 3, − | 3.5, 4, + | 23.8157 | 10 |
|  | 3, 1, 4, − | 3.5, 4, + | 23.8212 | -40 |
|  | 4, 1, 4, + | 4.5, 4, − | 23.8939 | 39 |
|  | 4, 1, 3, + | 4.5, 4, − | 23.8939 | -50 |
|  | 4, 1, 4, + | 4.5, 5, − | 23.9126 | 5 |
|  | 4, 1, 5, + | 4.5, 5, − | 23.9218 | 22 |
|  | 5, 1, 4, − | 5.5, 5, + | 24.0045 | 8 |
|  | 5, 1, 5, − | 5.5, 5, + | 24.0086 | 0 |
|  | 5, 1, 5, − | 5.5, 6, + | 24.0258 | 16 |
|  | 5, 1, 6, − | 5.5, 6, + | 24.0364 | 32 |
| $^Q Q_{10} + ^Q R_{10}$ | 0, 0, 0, + | 0.5, 1, − | 23.8472 | -46 |
|  | 1, 0, 1, − | 1.5, 1, + | 23.8816 | 51 |
|  | 1, 0, 1, − | 1.5, 2, + | 23.8989 | 7 |
|  | 2, 0, 2, + | 2.5, 2, − | 23.9472 | 14 |
|  | 2, 0, 2, + | 2.5, 3, − | 23.9647 | 11 |
|  | 3, 0, 3, − | 3.5, 3, + | 24.0294 | 15 |
|  | 3, 0, 3, − | 3.5, 4, + | 24.0466 | 23 |
| $^R R_{11}$ | 0, 1, 1, + | 1.5, 2, − | 25.0257 | 14 |
|  | 1, 1, 2, − | 2.5, 3, + | 26.0203 | -37 |
|  | 1, 1, 1, − | 2.5, 2, + | 26.0323 | 4 |
|  | 2, 1, 3, + | 3.5, 4, − | 27.0330 | -22 |
|  | 2, 1, 2, + | 3.5, 3, − | 27.0432 | 3 |
|  | 3, 1, 4, − | 4.5, 5, + | 28.0612 | -16 |
|  | 3, 1, 3, − | 4.5, 4, + | 28.0703 | 17 |
| $^R R_{10}$ | 0, 0, 0, + | 1.5, 1, − | 25.2687 | 46 |
|  | 1, 0, 1, − | 2.5, 2, + | 26.2617 | -21 |
|  | 2, 0, 1, + | 2.5, 3, − | 27.2721 | -39 |
| Std. dev. of fit: 27 MHz (0.00089 cm$^{-1}$); **(70 lines to 65 features)** | | | | |
|  |  |  |  |  |

**Table S3.** Spectroscopic parameters in wavenumbers (cm$^{-1}$) used to model the $\tilde{X}\,^2\Sigma^+(0,0,0)$ and $\tilde{A}\,^2\Pi_{1/2}(0,0,0)$ states of YbOH.

| | Par. | $^{170}$YbOH | $^{171}$YbOH[d] | $^{172}$YbOH[e] | $^{173}$YbOH[f] | $^{174}$YbOH[g] | $^{176}$YbOH |
|---|---|---|---|---|---|---|---|
| $X\,^2\Sigma^+$ | $B$ | 0.2456508 (fix)[c] | 0.245497(22) | 0.245387(6) | 0.245211(18) | 0.245116257(10) | 0.244846 (fix) |
| | $D\times 10^7$ | 2.689(fix) | 2.524 (fix) | 2.359[h] | 2.190(fix) | 2.029(13) | 1.699(fix) |
| | $\gamma$ | -0.002693 (fix) | -0.002697 (fix) | -0.00270(2) | -0.002704 (fix) | -0.0027068 (19) | -0.002714 (fix) |
| | $b_F$(Yb) | - | 0.22761(33) | - | -0.062817(67) | - | - |
| | $c$(Yb) | - | 0.0078(14) | - | -0.00273(45) | - | - |
| | $e^2Qq_0$(Yb) | - | - | - | -0.1107(16) | - | - |
| | $b_F$(H)[a] | 0.000160 (fix) | 0.000160 (fix) | 0.000160 (fix) | 0.000160 (fix) | 0.000160(60) | 0.000160 (fix) |
| | $c$(H)[a] | 0.000082 (fix) | 0.000082 (fix) | 0.000082 fix) | 0.000082 (fix) | 0.000082(16) | 0.000082 (fix) |
| | | | - | | | | |
| $A\,^2\Pi_{1/2}$ | $A$[b] | 1350.0 | 1350.0 | 1350.0 | 1350.0 | 1350.0 | 1350.0 |
| | $B$ | 0.253573 (fix) | 0.253435(24) | 0.253329(6) | 0.253185(16) | 0.253052(3) | 0.252808 (fix) |
| | $D\times 10^7$ | 2.709(fix) | 2.608(fix) | 2.506(fix) | 2.4052(fix) | 2.319[h] | 2.100(fix) |
| | $p+2q$ | -0.43895 (fix) | -0.438667 (82) | -0.43850(4) | -0.438457 (64) | -0.43807 (7) | -0.43760 (fix) |
| | $h_{1/2}$(Yb) | - | 0.0148(15) | - | -0.00422(20) | - | - |
| | $d$(Yb) | - | 0.03199(58) | - | -0.00873(13) | - | - |
| | $e^2Qq_0$(Yb) | - | - | - | -0.0642(17) | - | - |
| | $T_{00}$ | 17998.6502 (fix) | 17998.63619 (24) | 17998.61549 (9) | 17998.60268 (13) | 17998.5875(2) | 17998.5550 (fix) |

a) $\tilde{X}\,^2\Sigma^+(0,0,0)$ proton-hyperfine fixed to value determined for $^{174}$YbOH (Ref.36)

b) Spin-orbit parameter, *A*, fixed to values from analysis of high-temperature sample (Ref. 49)
c) "fix" values are obtained by extrapolation of the experimentally determined parameters using expected mass dependence.
d) $^{171}$YbOH parameters from present study.
e) $^{172}$YbOH parameters from Ref. 34.
f) $^{173}$YbOH parameters from present study.
g) $^{174}$YbOH $\tilde{X}\,^2\Sigma^+(0,0,0)$ values from analysis of microwave spectra (Ref.36 ); $^{174}$YbOH $\tilde{A}\,^2\Pi_{1/2}(0,0,0)$ values from Ref.36
h) *D* value from analysis of high temperature spectra (Ref. 49)

**Table S4**. The calculated low-rotational ($N$=0-4) energies for the $\tilde{X}\,^2\Sigma^+$ (0,0,0) state of $^{171}$YbOH.

| N | G | $F_1$ | F | Par. | Energy (cm$^{-1}$) | N | G | $F_1$ | F | Par. | Energy |
|---|---|---|---|---|---|---|---|---|---|---|---|
| 0 | 0 | 0 | 0.5 | + | 0.000000 | 3 | 0 | 3 | 2.5 | - | 2.945842 |
| 0 | 1 | 1 | 0.5 | + | 0.227523 | 3 | 0 | 3 | 3.5 | - | 2.945844 |
| 0 | 1 | 1 | 1.5 | + | 0.227644 | 3 | 1 | 4 | 3.5 | - | 3.169021 |
| 1 | 0 | 1 | 0.5 | - | 0.490978 | 3 | 1 | 4 | 4.5 | - | 3.169101 |
| 1 | 0 | 1 | 1.5 | - | 0.490979 | 3 | 1 | 3 | 2.5 | - | 3.176265 |
| 1 | 1 | 2 | 1.5 | - | 0.716935 | 3 | 1 | 3 | 3.5 | - | 3.176297 |
| 1 | 1 | 2 | 2.5 | - | 0.717029 | 3 | 1 | 2 | 2.5 | - | 3.177885 |
| 1 | 1 | 0 | 0.5 | - | 0.718704 | 3 | 1 | 2 | 1.5 | - | 3.177931 |
| 1 | 1 | 1 | 0.5 | - | 0.721210 | 4 | 0 | 4 | 3.5 | + | 4.909697 |
| 1 | 1 | 1 | 1.5 | - | 0.721289 | 4 | 0 | 4 | 4.5 | + | 4.909700 |
| 2 | 0 | 2 | 1.5 | + | 1.472930 | 4 | 1 | 5 | 4.5 | + | 5.131554 |
| 2 | 0 | 2 | 2.5 | + | 1.472932 | 4 | 1 | 5 | 5.5 | + | 5.131631 |
| 2 | 1 | 3 | 2.5 | + | 1.697467 | 4 | 1 | 4 | 3.5 | + | 5.140253 |
| 2 | 1 | 3 | 3.5 | + | 1.697552 | 4 | 1 | 4 | 4.5 | + | 5.140275 |
| 2 | 1 | 2 | 1.5 | + | 1.703217 | 4 | 1 | 3 | 3.5 | + | 5.143258 |
| 2 | 1 | 2 | 2.5 | + | 1.703294 | 4 | 1 | 3 | 2.5 | + | 5.143310 |
| 2 | 1 | 1 | 1.5 | + | 1.703356 | | | | | | |
| 2 | 1 | 1 | 0.5 | + | 1.703362 | | | | | | |
| | | | | | | | | | | | |
| | | | | | | | | | | | |

**Table S5**. The calculated low-rotational (*J*=0.5-5.5) energies for the $\tilde{A}\,^2\Pi_{1/2}(0,0,0)$ state of $^{171}$YbOH.

| J | $F_1$ | Par. | Energy (cm$^{-1}$)[a] | J | $F_1$ | Par. | Energy(cm$^{-1}$)[a] |
|---|---|---|---|---|---|---|---|
| 0.5 | 0 | - | 3.817599 | 3.5 | 3 | + | 6.974518 |
| 0.5 | 1 | - | 3.848790 | 3.5 | 4 | + | 6.991659 |
| 0.5 | 1 | + | 4.274554 | 3.5 | 4 | - | 8.731335 |
| 0.5 | 0 | + | 4.288257 | 3.5 | 4 | - | 8.746477 |
| 1.5 | 1 | + | 4.370924 | 4.5 | 4 | - | 9.035841 |
| 1.5 | 2 | + | 4.389693 | 4.5 | 5 | - | 9.052723 |
| 1.5 | 2 | - | 5.252854 | 4.5 | 5 | + | 11.230881 |
| 1.5 | 1 | - | 5.267347 | 4.5 | 4 | + | 11.246175 |
| 2.5 | 2 | - | 5.419697 | 5.5 | 5 | + | 11.603760 |
| 2.5 | 3 | - | 5.437311 | 5.5 | 6 | + | 11.620478 |
| 2.5 | 3 | + | 6.738611 | | | | |
| 2.5 | 2 | + | 6.753517 | | | | |

a) Term energies -17320.0 cm$^{-1}$

**Table S6**. The calculated low-rotational ($N$=0-4) energies for the $\tilde{X}\,^2\Sigma^+$ (0,0,0) state of $^{173}$YbOH.

| N | G | $F_1$ | F | Par. | Energy (cm$^{-1}$) | N | G | $F_1$ | F | Par. | Energy (cm$^{-1}$) |
|---|---|---|---|---|---|---|---|---|---|---|---|
| 0 | 3 | 3 | 2.5 | + | 0 | 3 | 3 | 6 | 6.5 | - | 2.948770 |
|   | 3 | 3 | 3.5 | + | 0.000094 |   | 3 | 2 | 1.5 | - | 2.954522 |
|   | 2 | 2 | 2.5 | + | 0.188530 |   | 3 | 2 | 2.5 | - | 2.954564 |
|   | 2 | 2 | 1.5 | + | 0.188603 |   | 3 | 1 | 0.5 | - | 2.965770 |
| 1 | 3 | 3 | 2.5 | - | 0.472924 |   | 3 | 1 | 1.5 | - | 2.965796 |
|   | 3 | 3 | 3.5 | - | 0.473026 |   | 3 | 0 | 0.5 | - | 2.972241 |
|   | 3 | 4 | 3.5 | - | 0.495229 |   | 3 | 6 | 5.5 | - | 2.948694 |
|   | 3 | 4 | 4.5 | - | 0.495313 | 3 | 2 | 4 | 4.5 | - | 3.121217 |
|   | 3 | 2 | 1.5 | - | 0.506711 |   | 2 | 4 | 3.5 | - | 3.121263 |
|   | 3 | 2 | 2.5 | - | 0.506790 |   | 2 | 3 | 3.5 | - | 3.122272 |
|   | 2 | 2 | 2.5 | - | 0.664003 |   | 2 | 3 | 2.5 | - | 3.122302 |
|   | 2 | 2 | 1.5 | - | 0.664066 |   | 2 | 2 | 2.5 | - | 3.132948 |
|   | 2 | 3 | 3.5 | - | 0.684417 |   | 2 | 2 | 1.5 | - | 3.132956 |
|   | 2 | 3 | 2.5 | - | 0.684476 |   | 2 | 5 | 5.5 | - | 3.141422 |
|   | 2 | 1 | 1.5 | - | 0.692711 |   | 2 | 5 | 4.5 | - | 3.141477 |
|   | 2 | 1 | 0.5 | - | 0.692767 |   | 2 | 1 | 0.5 | - | 3.144718 |
| 2 | 3 | 4 | 3.5 | + | 1.457907 |   | 2 | 1 | 1.5 | - | 3.144749 |
|   | 3 | 4 | 4.5 | + | 1.457993 | 4 | 3 | 5 | 4.5 | + | 4.892346 |
|   | 3 | 3 | 2.5 | + | 1.463750 |   | 3 | 5 | 5.5 | + | 4.892409 |
|   | 3 | 3 | 3.5 | + | 1.463831 |   | 3 | 6 | 5.5 | + | 4.893482 |
|   | 3 | 5 | 4.5 | + | 1.477345 |   | 3 | 6 | 6.5 | + | 4.893553 |
|   | 3 | 5 | 5.5 | + | 1.477424 |   | 3 | 4 | 3.5 | + | 4.899911 |
|   | 3 | 2 | 1.5 | + | 1.479419 |   | 3 | 4 | 4.5 | + | 4.899960 |
|   | 3 | 2 | 2.5 | + | 1.479491 |   | 3 | 7 | 6.5 | + | 4.909900 |
|   | 3 | 1 | 0.5 | + | 1.495662 |   | 3 | 7 | 7.5 | + | 4.909975 |
|   | 3 | 1 | 1.5 | + | 1.495731 |   | 3 | 3 | 2.5 | + | 4.911395 |

|   | 2 | 3 | 3.5 | + | 1.647900 |   | 3 | 3 | 3.5 | + | 4.911423 |
|---|---|---|-----|---|----------|---|---|---|-----|---|----------|
|   | 2 | 3 | 2.5 | + | 1.647952 |   | 3 | 2 | 1.5 | + | 4.923221 |
|   | 2 | 2 | 2.5 | + | 1.654388 |   | 3 | 2 | 2.5 | + | 4.923222 |
|   | 2 | 2 | 1.5 | + | 1.654428 |   | 3 | 1 | 1.5 | + | 4.932653 |
|   | 2 | 4 | 4.5 | + | 1.668139 |   | 3 | 1 | 0.5 | + | 4.932697 |
|   | 2 | 4 | 3.5 | + | 1.668195 |   | 2 | 4 | 4.5 | + | 5.082731 |
|   | 2 | 1 | 1.5 | + | 1.668491 |   | 2 | 4 | 3.5 | + | 5.082757 |
|   | 2 | 1 | 0.5 | + | 1.668516 |   | 2 | 5 | 5.5 | + | 5.084402 |
|   | 2 | 0 | 0.5 | + | 1.678606 |   | 2 | 5 | 4.5 | + | 5.084446 |
| 3 | 3 | 5 | 4.5 | - | 2.931020 |   | 2 | 3 | 3.5 | + | 5.091379 |
|   | 3 | 5 | 5.5 | - | 2.931099 |   | 2 | 3 | 2.5 | + | 5.091380 |
|   | 3 | 4 | 3.5 | - | 2.931727 |   | 2 | 2 | 1.5 | + | 5.103627 |
|   | 3 | 4 | 4.5 | - | 2.931800 |   | 2 | 2 | 2.5 | + | 5.103661 |
|   | 3 | 3 | 2.5 | - | 2.941655 |   | 2 | 6 | 6.5 |   | 5.104719 |
|   | 3 | 3 | 3.5 | - | 2.941713 |   | 2 | 6 | 5.5 |   | 5.104773 |

**Table S7**. The calculated low-rotational (*J*=0.5-5.5) energies for the $\tilde{A}\,^2\Pi_{1/2}(0,0,0)$ state of $^{173}$YbOH.

| J | $F_1$ | Par. | Energy (cm$^{-1}$)[a] | J | $F_1$ | Par. | Energy |
|---|---|---|---|---|---|---|---|
| 0.5 | 3 | − | 3.704644 | 3.5 | 4 | − | 8.599705 |
|  | 2 | − | 3.730533 |  | 3 | − | 8.600887 |
| 0.5 | 2 | + | 4.148686 |  | 2 | − | 8.604951 |
|  | 3 | + | 4.154094 |  | 5 | − | 8.605094 |
| 1.5 | 4 | + | 4.248032 |  | 1 | − | 8.609171 |
|  | 3 | + | 4.252934 |  | 6 | − | 8.621879 |
|  | 2 | + | 4.263804 | 4.5 | 6 | − | 8.905220 |
|  | 1 | + | 4.279865 |  | 7 | − | 8.908700 |
| 1.5 | 3 | − | 5.124399 |  | 5 | − | 8.909227 |
|  | 2 | − | 5.125466 |  | 4 | − | 8.917751 |
|  | 1 | − | 5.132311 |  | 3 | − | 8.928048 |
|  | 4 | − | 5.140742 |  | 2 | − | 8.937828 |
| 2.5 | 4 | − | 5.295207 | 4.5 | 5 | + | 11.097502 |
|  | 5 | − | 5.295846 |  | 4 | + | 11.097569 |
|  | 3 | − | 5.302368 |  | 3 | + | 11.101110 |
|  | 2 | − | 5.313638 |  | 6 | + | 11.103870 |
|  | 1 | − | 5.324084 |  | 2 | + | 11.105793 |
|  | 0 | − | 5.330200 |  | 7 | + | 11.120350 |
| 2.5 | 3 | + | 6.608434 | 5.5 | 7 | + | 11.470184 |
|  | 2 | + | 6.611476 |  | 6 | + | 11.473392 |
|  | 4 | + | 6.611959 |  | 8 | + | 11.474155 |
|  | 1 | + | 6.616200 |  | 5 | + | 11.481249 |
|  | 0 | + | 6.619376 |  | 4 | + | 11.491456 |
|  | 5 | + | 6.628960 |  | 3 | + | 11.502043 |
| 3.5 | 5 | + | 6.846715 |  |  |  |  |
|  | 6 | + | 6.849298 |  |  |  |  |
|  | 4 | + | 6.851926 |  |  |  |  |
|  | 3 | + | 6.861448 |  |  |  |  |
|  | 2 | + | 6.871831 |  |  |  |  |
|  | 1 | + | 6.880325 |  |  |  |  |

a) Term energies -17320.0 cm$^{-1}$